\documentclass[3p]{elsarticle}

\usepackage[T1]{fontenc}
\usepackage[utf8]{inputenc}
\usepackage[english]{babel}
\usepackage{float}
\usepackage{amsmath}
\usepackage{wasysym}
\usepackage{nicefrac}
\usepackage{mathtools}
\usepackage{amssymb}
\usepackage{mathtools}
\usepackage{stmaryrd}
\providecommand{\abs}[1]{\lvert#1\rvert}

\newcommand\eqdef{\mathrel{\overset{{{\tiny\mathrm{def}}}}{=}}}

\usepackage{tikz}
\usepackage{todonotes}
\usepackage{csquotes}
\usepackage{booktabs}
\usepackage{makecell}
\usepackage{siunitx}
\usepackage{dcolumn}
\usepackage{multirow}
\usepackage{listings}
\usepackage{soul}
\usepackage{placeins}
\usepackage{tabularx} 
\usepackage[vlines]{tabularht}
\usepackage{adjustbox}
\usepackage{makecell}
\usepackage{fullpage}
\usepackage{threeparttable}
\usepackage{wasysym}
\usepackage{diagbox}
\usepackage{multirow}
\usepackage{caption}
\usepackage{subcaption}
\usepackage[figuresright]{rotating}
\usepackage{xspace}
\usepackage{comment}

\usepackage{algorithm}
\usepackage{algpseudocode}

\providecommand\longrightarrowRHD{\relbar\joinrel\relbar\joinrel\mathrel\RHD}
\providecommand\longrightarrowrhd{\relbar\joinrel\relbar\joinrel\mathrel\rhd}

\makeatletter
\providecommand*\xrightarrowRHD[2][]{\ext@arrow 0055{\arrowfill@\relbar\relbar\longrightarrowRHD}{#1}{#2}}
\providecommand*\xrightarrowrhd[2][]{\ext@arrow 0055{\arrowfill@\relbar\relbar\longrightarrowrhd}{#1}{#2}}
\makeatother

\usepackage{paralist}

\usepackage{amsthm}
\newtheorem{definition}{Definition}%
\newtheorem{lemma}{Lemma}%
\newtheorem{proposition}{Proposition}%
\newtheorem{claim}{Claim}%

\newdefinition{rmk}{Remark}

\usepackage{lineno}
\modulolinenumbers[0]
\usepackage[obeyspaces,hyphens]{url}
\PassOptionsToPackage{hyphens}{url}
\usepackage[]{hyperref}
\hypersetup{
	colorlinks=true,
	pdftex,
	pdfauthor={Dominik Grzelak},
	pdftitle={Generating random bigraphs with preferential attachment},
	pdfkeywords={bigraphs, random graphs, preferential attachment, assortativity, average neighbor degree, ubiquitous systems}
}

\usepackage[capitalise,noabbrev]{cleveref}
\Crefname{figure}{Fig.}{Figs.}%
\Crefname{section}{Sec.}{Sec.}%
\Crefname{equation}{Eq.}{Eq.}%
\Crefname{definition}{Def.}{Defs.}%
\Crefname{table}{Tab.}{Tabs.}%
\Crefname{lemma}{Lemma}{Lemma}%

\journal{SI on Discrete Models of Complex Systems: recent trends and analytical challenges}

\begin{document}

\begin{frontmatter}
	
	\title{Generating random bigraphs with preferential attachment}
	
	\author[tudaddress]{Dominik Grzelak\corref{mycorrespondingauthor}\fnref{myfootnote}}
	\ead{dominik.grzelak@tu-dresden.de}
	
	\author[reutlingendaddress]{Barbara Priwitzer}
	\ead{barbara.priwitzer@reutlingen-university.de}

	\author[tudaddress]{Uwe Aßmann\fnref{myfootnote}}
	\ead{uwe.assmann@tu-dresden.de}

	\address[tudaddress]{Software Technology Group, Technische Universität Dresden, Germany}
	\address[reutlingendaddress]{Fakultät Technik, Hochschule Reutlingen, Germany}
	\cortext[mycorrespondingauthor]{Corresponding author}
	\fntext[myfootnote]{Dominik Grzelak and Uwe A{\ss}mann are also with Centre for Tactile Internet with Human-in-the-Loop (CeTI), Technische Universität Dresden, 01062 Dresden, Germany.}
	
	\begin{abstract}
		The bigraph theory is a relatively young, yet formally rigorous, mathematical framework encompassing Robin Milner's previous work on process calculi, on the one hand, and provides a generic \textit{meta-model} for \textit{complex systems} such as \textit{multi-agent systems}, on the other. 
		A bigraph $F = \langle F^P, F^L\rangle$ is a superposition of two independent graph structures comprising a place graph $F^P$ (i.e., a forest) and a link graph $F^L$ (i.e., a hypergraph), sharing the same node set, to express locality and communication of processes independently from each other.
		
		In this paper, we take some preparatory steps towards an algorithm for generating random bigraphs with preferential attachment feature w.r.t. $F^P$ and assortative (disassortative) linkage pattern w.r.t. $F^L$. We employ parameters allowing one to fine-tune the characteristics of the generated bigraph structures. To study the pattern formation properties of our algorithmic model, we analyze several metrics from graph theory based on artificially created bigraphs under different configurations.
		
		Bigraphs provide a quite useful and expressive semantic for process calculi for mobile and global ubiquitous computing. So far, this subject has not received attention in the bigraph-related scientific literature. 
		However, artificial models may be particularly useful for simulation and evaluation of real-world applications in ubiquitous systems necessitating random structures.
	\end{abstract}
	
	
\end{frontmatter}

\section{Introduction and Motivation}
Complex systems exhibit highly intertwined agents (physical or logical) organized in more or less hierarchical-like structures where agents possess loose and close inter-linkages which additionally are equipped with different semantics (e.g., communication).
Consequently, collections of agents (so-called ensembles) lead to unpredictable collective behavior (cf. \cite[p.~2]{loke_representing_2016}). One can observe that the system's dynamic and emergent complexity is greater than the sum of the individual collections and their parts.
Hence, the development of any complex system is not feasible without understanding and responding to the inter-dependencies and linkages equally on the macro- and microscopic scale.

In this respect, \textit{agent-based models} have become of increasing importance in different fields of application regarding the modeling and simulation of complex systems (see, for example, \cite{heppenstall_agent-based_2012,namatame_agent-based_2016, cho_churning_2016, muller_abm_2017}). Within this computational model, individual entities of a complex system are modeled as \textit{agents}, whereas the behavior and interaction between them are determined by a set of predefined \textit{rules}. 
This rudimentary description coincides with the most common definition of agent-based models in the literature, namely, that agents act with each other within an environment over time in order to study systems exhibiting complex behavior (see \cite{crooks_introduction_2012} and \cite{metz_agent-based_2017} for a more detailed discussion).
Owing to the considerable success agent-based modeling approaches have enjoyed so far in various domains, many formalisms have emerged around that subject, among them different \textit{process calculi}.
Here, the term \textit{process} "refers to behavior of a system" which is, for example, the action of a user or the execution of a software system \cite[p.~19-1]{baeten_process_2007}.
Agents interact with each other through communication links, which are identified by \textit{names}. More specifically, the computational notion of such an algebra "is represented purely as the communication of names across links" \cite[p.~1]{milner_calculus_1992}.
Many process calculi provide a sufficient level of abstraction, and share key properties such as compositional modeling and behavioral reasoning via equivalences and preorders (see \cite[Ch.~1]{process_algebra_handbook_2001}), thus making them a remarkably straightforward and rich modeling language.
 
One variant of process calculi, we continue to focus in this paper, are \textit{bigraphs} devised by Robin Milner and colleagues (see, among others, \cite{milner_axioms_2004,jensen_bigraphs_2004,milner_space_2009}). Bigraphs are a relatively new model for interactive and distributed agents, where graphs are treated as the primary mathematical objects \cite[p.~16]{milner_bigraphical_2001}. Category theory is the broader underlying mathematical framework for axiomatizing and expressing bigraphs and their respective operations \cite[p.~14]{milner_space_2009}. The formal underpinning empowers a generic formalization of families of similar models and dynamical systems, which we unveil shortly.
In particular, the theory is dedicated to the following two principal aims, which can be identified in the scientific literature: 
\begin{enumerate}[i)] %
	\item Creating a unifying theory for several existing process-calculi frameworks (see \cite{milner_bigraphical_2001}). The particularly expressive bigraph structure enables the unification of a great variety of process calculi that focus on communication and locality \cite{krivine_stochastic_2008} such as Calculus of Communicating Systems (CSS) \cite{milner_calculus_1982}, $\pi$-calculus \cite{milner_communicating_1999} and mobile ambients \cite{cardelli_mobile_2000}. 
	Also bigraphical encodings for condition-event Petri nets have been studied in \cite{milner_space_2009,milner_bigraphs_2004-1, leifer_transition_2006}, capturing their syntax and semantic.
	Recently, the authors of the Spider Calculus \cite{pierce_spider_2010} indicated that bigraphs could represent this very calculus as well.
	Having a general meta-model for process calculi at hand that "can describe several
	concrete calculi" is a great benefit because "one can hope that a result for a meta-model can be transferred to all of these calculi" \cite[pp.~127-128]{elsborg_type_2009}.
	\item Developing a generic structural meta-model that enables modeling of ubiquitous systems (see \cite{milner_space_2009}). 
	So far, the bigraphical theory found application in various scientific fields where we wish to mention a few, particularly important ones.
	With respect to biology, Krivine et al. \cite{krivine_stochastic_2008} developed \textit{stochastic semantics for bigraphs} by using the process of membrane budding as an example, showing that bigraphs are a well-suited candidate for representing complex bio-molecular reactions.
	Damgaard and Krivine \cite{c._damgaard_generic_2008} further investigated the development of a generic language by using the "bigraphical framework as a basis for developing families of calculi for modelling of biological systems at the molecular level" \cite[p.~2]{c._damgaard_generic_2008}.
	An orthogonal work of the previous ones is \cite{bacci_bigraphical_2009} where \textit{biological bigraphs}, a meta-model at the protein-level, is developed.
	Considering the wide field of computer science, Birkedal et al. \cite{birkedal_bigraphical_2006} proposed the \textit{plato-graphical model} for the formal modeling of \textit{context-aware systems}.
	Particularly useful for ubiquitous systems, Sevegnani and coworkers \cite{sevegnani_bigraphs_2012, sevegnani_bigraphs_2015} proposed a generalization of Milner's bigraph theory called \textit{bigraph with sharing} which can model overlapping or intersecting realms (e.g., to model a shared space among many entities or overlapping wireless signals). Therefore, Sevegnani implemented directed acyclic graphs (DAG) instead of using trees for the place graph (see \Cref{sec:bigraphs}).
	Modeling \textit{socio-technical systems} is explored by Benford et al. \cite{benford_lions_2016} by means of the pervasive outdoor game called \textit{Savannah} \cite{benford_life_2005}, where a socio-technical system is understood as a facet of ubiquitous computing systems. The presented model includes four perspectives: a computational, physical, human-related, and technology-related one. Based on this, the authors demonstrate the analysis of complex interactional phenomena and exploration of possible inconsistencies among the four perspectives of this formal model \cite{benford_lions_2016}.
\end{enumerate}

Related to the second mentioned long-term objective of Milner, we wish to investigate on the random generation of bigraphs. 
Random bigraph generation is especially useful in applications necessitating the usage of random structures and for simulations of real-world systems (see also \cite{nobari_fast_2011}).
Another compelling reason for creating synthetic bigraphs includes \textit{benchmarking} of bigraph-related algorithms.
In order to foster advances and evaluation of different algorithms such as \textit{frequent bigraph mining} (which has yet to be developed, analogously to frequent subtree mining \cite{chi_frequent_2005}) or \textit{bigraph matching}, the development of bigraph databases are inevitable.
Synthetic graphs can then be acquired from such a bigraph database and derivatives constructed easily, ready to be used for a series of experiments. For example, Bunke et al. \cite{bunke_matching_2008} used a graph database from \cite{foggia_database_2001}\footnote{The database was designed to assess graph and subgraph isomorphism problems (see also \cite{de_santo_large_2003} for further reference.). It is, however, not available anymore at the URL \url{http://amalfi.dis.unina.it/graph}.} to study the computational complexity of their hypergraph matching algorithm.
We can observe that such graph databases are a valuable resource and that it is worth to follow this issue much further with respect to bigraphs. 

\subsection{Related Work}

To the best of our knowledge, random bigraph generation was not the main focus of related scientific research until now but extensively for similar graphical structures, e.g., DAGs, networks, trees, and forests, which we are briefly present below.

A great many random network models have been studied so far, including the configuration model, the Erdös-Rényi (ER) model \cite{erdos_random_1959} and the Watts-Strogatz model (i.e., a small-world network model) \cite{watts_collective_1998}. We refer the reader to \cite{barabasi_network_2016} for a greater overview. 
This list is not exhaustive, but worth mentioning are further Wilson's algorithm  \cite{wilson_generating_1996} for creating random spanning trees of an undirected graph, dynamic network models \cite{zhang_random_2017} or random graphs with fixed degree sequences \cite{fosdick_configuring_2018}. Most of them are sufficiently matured, evaluated, and tested, and found their practical applicability in concrete domains.
On the contrary, observed from a pragmatic \textit{model transformation} standpoint, Fernández et al. \cite{fernandez_labelled_2016} present a different approach by means of graph rewriting. A \textit{port graph} is used as the internal embodiment of a social network. In this social network model, nodes represent people and edges the connections between them. The graph is then incrementally expanded by consecutively applying rules which reconfigure the social network graph.

\subsection{Contributions of this Paper}

In this work, we shall focus on taking some preparatory steps towards an algorithm for generating random bigraphs by extending the ideas of previous works. Due to the bigraph's anatomy, however, former ideas are not always directly applicable within the framework of bigraphs. Thus, the generation differs from the usual network generation approaches and must be appropriately adapted. 
In particular, the main contribution of this paper is to investigate on the construction of \textit{bigraphical agents}, a certain kind of bigraphs that are being appropriated by bigraphical reactive systems (BRS)\footnote{BRSs are an extension for bigraphs, covering the dynamic aspect of the theory (see \cref{sec:bigraphs-dynamic}).}.
The reason for this: Reduction semantics in process calculi are commonly prescribed by rules (i.e., reactions) of the form $a \rightarrowtriangle a'$, where $a, a'$ are called \textit{agents} (see also \cite{milner_space_2009,milner_bigraphical_2001}); making them also one of the main algebraic structures concerning bigraph matching and rewriting (see \Cref{sec:bigraphs-dynamic}). Both are necessarily the primary operations to cover the computational notion of bigraphs. Meaning, they are used almost without exception when conducting simulations or employing the bigraph theory for real-world applications.

\subsection{Structure of this Paper}
The paper is structured as follows.
In \cref{sec:bigraphs}, we begin by introducing bigraphs and bigraphical reactive systems. Then in \cref{sec:algorithm}, we present our approach for random generation of place graphs first, and after link graphs. In \cref{sec:analysis}, we compute several measures originating from network theory and conduct statistical analyses to capture information about the structure of the generated bigraphs and the algorithm's behavior. For example, we use global measures such as the assortativity for analyzing link graphs and focus on the node degree distribution of place graphs. 
We provide proof outlines of some properties based on statistical analyses of experimental data. Finally, \cref{sec:conclusion} concludes this paper and gives some directions for future work.

\section{Preliminaries: Bigraphs and Bigraphical Reactive Systems}\label{sec:bigraphs}

\begin{figure}[thbp]
	\centering
	\includegraphics[width=0.85\linewidth]{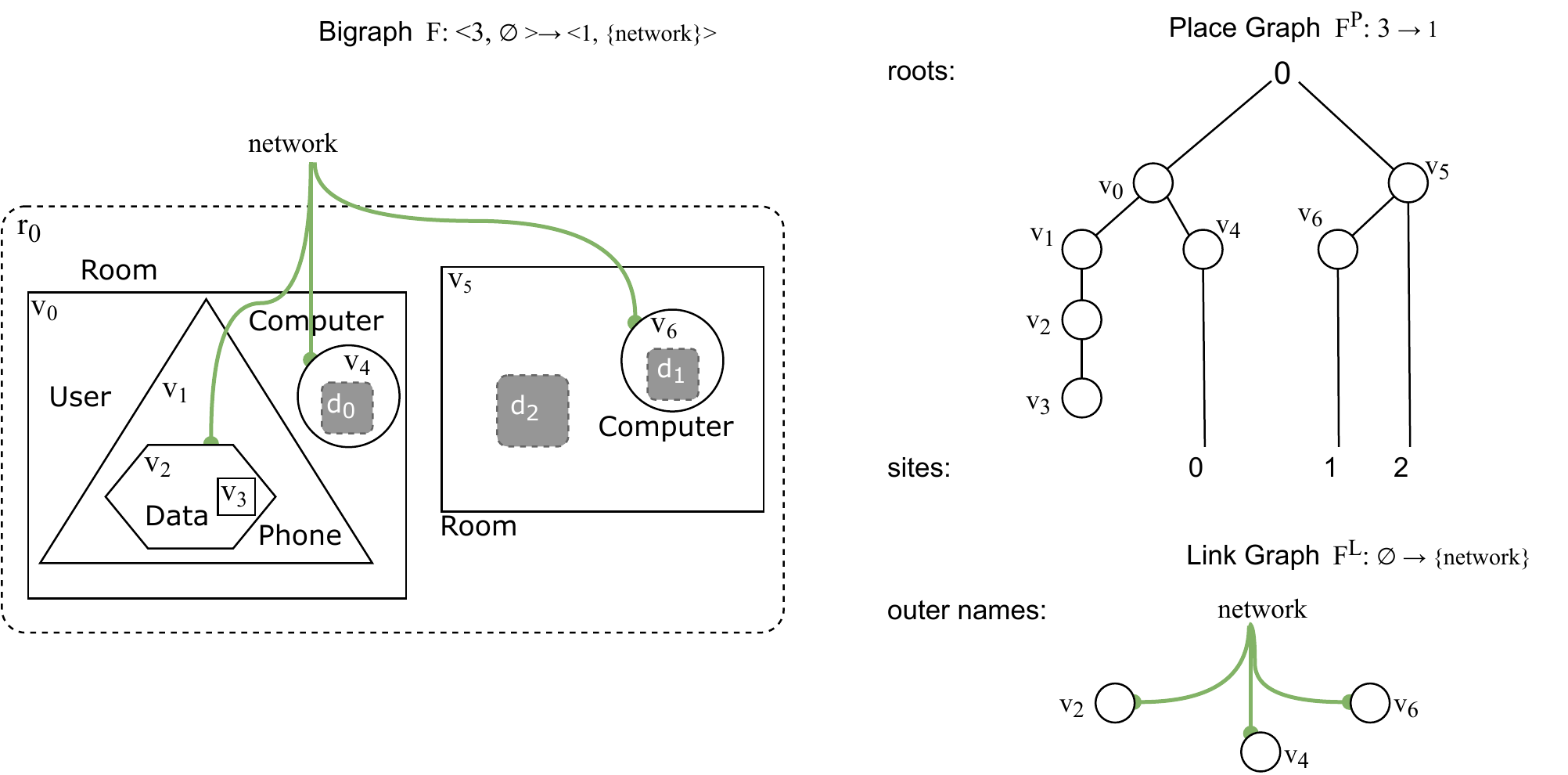}
	\caption{A concrete bigraph with its corresponding place graph and link graph. The bigraph has one outer name $Y=\{network\}$, and one root $n=\{0\}$. It has three sites $m=\{0,1,2\}$ and no inner names $X=\emptyset$.}
	\label{fig:bigraph-definition}
\end{figure}

We wish to give a brief introduction concerning the bigraph theory devised by Robin Milner. In the sequel, we use the formalization based on \cite{milner_space_2009} for this purpose, where the reader can find all further relevant formal definitions. A bigraph is either \textit{abstract} or \textit{concrete}. Specifically in this work, we are concerned with \textit{pure} concrete bigraphs (see also \cite{milner_space_2009,milner_axioms_2005,milner_pure_2006}). Bigraphs employ category theory as the formal mathematical underpinning; however, the reader does not need to be concerned about technical details at this point.
For the following explanation of the bigraph's anatomy, we shall use the bigraph depicted in \cref{fig:bigraph-definition} as a running example.

\paragraph{Signature}
A bigraph is defined over a signature $\mathcal{K}$ which specifies the syntax of the bigraph. Let $C = \{C_i\}_{0 \leq i \leq n-1}$ denote the set of controls of a bigraph $B$ over the signature $\mathcal{K}$, $C \in \mathcal{K}$. The signature of the bigraph $F$ in \cref{fig:bigraph-definition} is $\mathcal{K} = \{\mathsf{Room}\colon 0, \mathsf{Computer}\colon 1, \mathsf{User}\colon 0, \mathsf{Phone}\colon 1, \mathsf{Data}\colon 0\}$.

\begin{definition}{(Basic Signature (after \cite[p.~7]{milner_space_2009}))}\label{def:signature}
	A basic signature takes the form $(\mathcal{K}, ar)$. It has a set $\mathcal{K}$ whose elements are kinds of node called \textit{controls}, and a map $ar\colon \mathcal{K} \to \mathbb{N}$ assigning an \textit{arity}, a natural number, to each control. The signature is denoted by $\mathcal{K}$ when the arity is understood. A bigraph over $\mathcal{K}$ assigns to each node a control, whose arity indexes the \textit{ports} of a node, where links may be connected.
\end{definition}

\subsection{Static Structure of a Bigraph}\label{sec:bigraphs-static}
Having arrived here, we define place graphs first, and link graphs after to finally combine them into bigraphs. Both graph structures are the constituents of a bigraph which are independently defined and only share the same node set.

\paragraph{Place Graph} A place graph is a forest of trees with special leaves. A tree of a place graph is a rooted unordered labeled graph, which is acyclic. 
It is used to model node hierarchies that reflect the parent-child-relations and is represented visually by nesting nodes into other nodes. Thus, inner nodes contained in outer nodes are called children of these. For instance, node $v_1$ with $ctrl(v_1) = \mathsf{User}$ is a child of $v_0$ with $ctrl(v_0) = \mathsf{Room}$. This nesting is used to express physical or logical location, ownership, or similar concepts (referring to \cref{fig:bigraph-definition}, a room contains a computer and a user, where a user has a phone). \textit{Siblings} are nodes under the same parent.
The gray shaded shapes are special leaves of the place graph, called \textit{sites}, and represent "holes". They have two purposes. On the one hand, they abstract other nodes away, and on the other, they are used to nest other bigraph's roots in these sites. A \textit{root} of a place graph is also called "region" (illustrated as an outermost container with a dotted border). Distinct finite ordinals index both roots and sites and represent the outer face and inner face of the place graph, respectively. %

\begin{definition}{(Concrete Place Graph (after \cite[p.~15]{milner_space_2009}))}\label{def:placeGraph}
A \textit{concrete place graph}
\begin{equation*}
	F = (V_F, ctrl_F, prnt_F)\colon m \to n
\end{equation*}
is a triple having an inner face $m$ and an outer face $n$, both finite ordinals. These index respectively the \textit{sites} and \textit{roots} of the place graph. $F$ has a finite set $V_F \subset \mathcal{V}$ of nodes, a control map $ctrl_F\colon V_F \to \mathcal{K}$ and a \textit{parent map} $prnt_F\colon m \uplus V_F \to V_F \uplus n$ which is acyclic, i.e. if $prnt_{F}^{i}(v) = v$ for some $v \in V_F$ then $i=0$.
\end{definition}

\paragraph{Link Graph} A link graph is a hypergraph where edges also connect to "special" graph elements that form the interface of the link graph. These are called \textit{inner names} and \textit{outer names}.
An integral part of a node is the port (see also \cref{def:signature}). The arity of a control determines how many ports a node has, meaning, how many links can be connected to this node. Each distinct node's port can directly connect to an edge or an outer name. Links connect multiple \textit{points}, i.e., \textit{inner names} and \textit{ports}. This notion is expressed by the link graph depicted at the bottom-right of \cref{fig:bigraph-definition}.

\begin{definition}{(Concrete Link Graph (after \cite[p.~15]{milner_space_2009}))}\label{def:linkGraph}
A \textit{concrete link graph}
\begin{equation*}
	F = (V_F, E_F, ctrl_F, link_F)\colon X \to Y
\end{equation*}
is a quadruple having an inner face $X$ and an outer face $Y$, both finite subsets of $\mathcal{X}$, called respectively the \textit{inner} and \textit{outer names} of the link graph. $F$ has finite sets  $V_F \subset \mathcal{V}$ of \textit{nodes} and $E_F \subset \mathcal{E}$ of \textit{edges}, a \textit{control map} $ctrl_F\colon V_F \to \mathcal{K}$ and a \textit{link map} $link_F\colon X \uplus P_F \to E_F \uplus Y$ where $P_F \stackrel{def}{=} \{\, (v,i) \mid i \in ar(ctrl_F(v)) \,\}$ is the set of \textit{ports} of $F$. Thus $(v,i)$ is the $i$th port of node $v$. We shall call $X \uplus P_F$ the \textit{points} of $F$, and $E_F \uplus Y$ its \textit{links}.
\end{definition}

Further, we distinguish between \textit{open} and \textit{closed} links. An outer name is an \textit{open link} and an edge is a \textit{closed link} \cite[p.~1013]{milner_axioms_2005}, and a point is open if its link is open, otherwise it is closed. A closed link can be thought of a "restricted" name which is invisible to the context and has no name attached to it, unlike an open name (see \cite[p.~129]{elsborg_type_2009}). A link is called \textit{idle} when no point is connected to it.
Referring to \cref{fig:bigraph-definition}, all devices such as the two computers in the different rooms and the user's phone are connected with the same network denoted by the outer name $network$ (i.e., they share the same \textit{link}).

Having defined a place graph and a link graph formally, we arrive at the definition for a concrete bigraph.

\begin{definition}{(Concrete Bigraph (see \cite[p.~15]{milner_space_2009}))}\label{def:bigraph}
A \textit{concrete bigraph} 
\begin{equation*}
F = (V_F, E_F, ctrl_F, prnt_F, link_F)\colon \langle k,X \rangle \to \langle m,Y \rangle
\end{equation*}
is a quintuplet comprising a concrete place graph $F^P = (V_F, ctrl_F, prnt_F)\colon k \to m $ and a concrete link graph $F^L = (V_F, E_F, ctrl_F, link_F)\colon X \to Y$. A concrete bigraph is also written as $F = \langle F^P, F^L\rangle$.
\end{definition}

\paragraph{Elementary Bigraphs} Elementary bigraphs are node-free bigraphs; an exceptions are \textit{ions}, \textit{atoms} and \textit{molecules}. We distinguish between \textit{placings} and \textit{linkings} (see \cite{milner_space_2009}).  

They represent the set of basic bigraphs of which more complex ones can be built. \Cref{tab:overview-elementary-bigraphs} shows some basic bigraphs in their graphical and corresponding algebraic notation.
\begin{table}[tbph]
	\centering
	\caption{Overview of some elementary bigraphs.}\label{tab:overview-elementary-bigraphs}
	\begin{tabular*}{\textwidth}{c @{\extracolsep{\fill}} cc}  %
		\toprule
		 & Notation & Example\\
		\midrule
	\parbox[t]{0mm}{\multirow{4}{*}{\rotatebox[origin=c]{90}{Placings $\phi$\qquad\quad}}} & $1\colon 0 \rightarrow 1$ & 1 = \raisebox{-0.6\totalheight}{\includegraphics[width=0.04\textwidth]{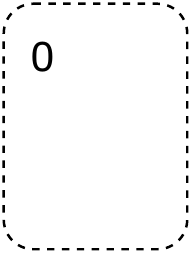}} \\ 
		& $join\colon 2 \rightarrow 1$ & $join =$ \raisebox{-0.6\totalheight}{\includegraphics[width=0.1\textwidth]{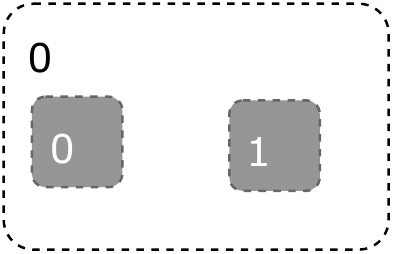}} \\ 
		& $\gamma_{1,1}\colon 2 \rightarrow 2$ & $\gamma_{1,1} =$   \raisebox{-0.6\totalheight}{\includegraphics[width=0.1\textwidth]{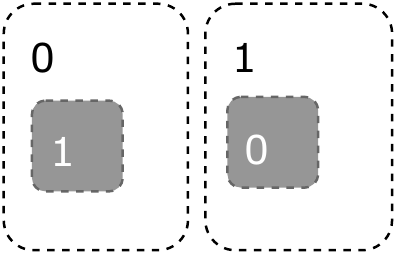}}\\ 
		& ${merge}_{n}\colon n \rightarrow 1$ & ${merge}_3 = $ \raisebox{-0.6\totalheight}{\includegraphics[width=0.1\textwidth]{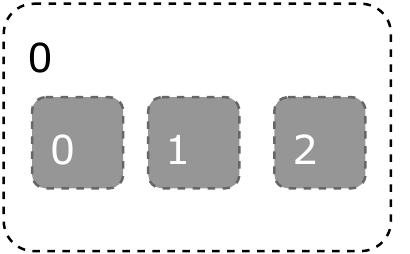}}\\
		 
			\midrule
			
\parbox[t]{0mm}{\multirow{2}{*}{\rotatebox[origin=c]{90}{Linkings $\lambda$}}} & elementary substitution $\nicefrac{y}{X}\colon X \rightarrow y$ & 
$\nicefrac{y}{\{x_1,x_2,...,x_n\}} = $ \raisebox{-0.50\totalheight}{\includegraphics[width=0.15\textwidth]{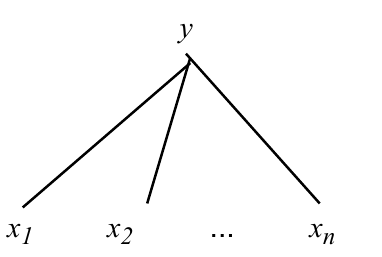}} \\ 
		
		& elementary closure ${/x}\colon x \rightarrow \epsilon$  & $/\{x_1,x_2,...,x_n\} =$ \raisebox{-0.6\totalheight}{\includegraphics[width=0.2\textwidth]{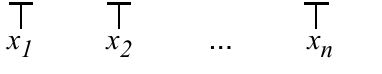}}\\ 
		\bottomrule 
	\end{tabular*} 
\end{table}
A bijection from sites to roots is called a \textit{permutation} $\pi$ and a bijective substitution is a \textit{renaming} $\alpha$.
A \textit{substitution} $\sigma\colon X \rightarrow Y$ is the tensor product of elementary substitutions $\sigma \eqdef \nicefrac{y_0}{X_0} \otimes \cdots \otimes \nicefrac{y_{n-1}}{X_{n-1}}$, where $Y = \{\vec{y}\}$ and $X_0 \uplus \cdots \uplus X_{n-1}$, and a \textit{closure} is the tensor product of elementary closures $/W \eqdef /w_0 \otimes \cdots \otimes /w_{n-1}$, where $W = \{\vec{w}\}$. The placing $merge$ itself is recursively defined ${merge}_0 = 1, {merge}_1 = \mathsf{id}_1, {merge}_2 = join, merge_{n+1} = join \circ (\mathsf{id}_1 \otimes {merge}_n)$ using only the identity place graph at $1$ and $join$.

\paragraph{Composition}

To conclude this section, we discuss some of the basic bigraph operations for constructing more sophisticated graph structures.
All bigraphs can be obtained from elementary bigraphs using the basic categorical operations, i.e., tensor product $\otimes$ for the juxtaposition of bigraphs and composition $\circ$ for nesting of bigraphs. Above, we have shown how to produce any substitution and closure by using just the tensor product. We shall illustrate their use by means of our running example bigraph $B$ over the signature $\mathcal{K}$ from \Cref{fig:bigraph-definition}. The full algebraic expression is
\begin{align*}
	B &= (\nicefrac{network}{\{a,b\}} \otimes join) \circ (A_1 \otimes A_0) \\
	& \qquad \qquad \qquad \mathrm{with} \\
	A_0 &= ( \mathsf{id}_{a} \otimes (\mathsf{Room} \circ join)) \circ (\mathsf{id}_1 \otimes \mathsf{Computer}_{a}) \\
	& \qquad \qquad \qquad \mathrm{~and} \\
	A_1 &= (\nicefrac{b}{\{x,y\}} \otimes \mathsf{Room} \circ join) \circ (((\nicefrac{y}{\{x\}} \otimes \mathsf{User}) \circ  \mathsf{Phone}_{x} \circ \mathsf{Data} \circ 1) \otimes \mathsf{Computer}_{x}). \\
\end{align*}

What we deliberately have left out, are derived operations that generalize the composition and tensor product, namely, \textit{nesting} "$\mathsf{.}$" and \textit{parallel product} "$||$", respectively, which allow name sharing. Using these derived operators yield in a more compact and convenient expression as the one presented above. For more information on this matter, we refer to \cite{milner_space_2009}.

\subsection{Dynamics of a Bigraph}\label{sec:bigraphs-dynamic}

For process calculi, it is common to express dynamics through reactions.
A reaction is a labeled transition of the form $a \xrightarrowrhd{L} a'$ where $\xrightarrowrhd{L}$ is regarded as a reaction relation.
Comparing two systems, whether they behave alike, is primarily conducted by means of a labeled transition system (LTS). In general, an LTS is a semantic model to describe distributed systems, which can be thought of as a directed graph with nodes and edges. Nodes are called the states of the system, and edges represent the transitions. Each edge is labeled from some vocabulary expressing the action between two states. Thus, the label $L$ indicates the action on how to reach the specific state $a'$ from $a$.

To express dynamics within bigraphs, the extension called bigraphical reactive systems was introduced by Milner.
Different notions of behavioral equivalence exist, such as trace equivalence or bisimilarity. The bigraph theory provides a uniformly treat across process calculi and a proof that bisimilarity is always a congruence in a \textit{wide reactive system} with RPOs \cite[Theorem~7.16]{milner_space_2009}.
Here we briefly present the technical definitions. Readers interested in a more comprehensive presentation of this topic may consult \cite{jensen_bigraphs_2004,milner_space_2009,jensen_bigraphs_2003}.
In contrast to the reaction relation mentioned above $a \xrightarrowrhd{L} a'$, the key question here is whether these labels can be derived from a set of reaction rules of the form $r \longrightarrowrhd r'$ instead. Within the bigraphical framework, labels are regarded as \textit{contexts} so that $a \xrightarrowrhd{L} a'$ implies the reaction $L \circ a \longrightarrowrhd a'$. \cite[p.~18]{milner_bigraphical_2001}
Here, the label represents how an "environment" contributes to the transition such that $L \circ r \longrightarrowrhd L \circ r'$ and $a' = L \circ r'$ being minimal w.r.t. $L$ for a set of reaction rules. \cite[p.~17]{milner_bigraphical_2001-1}

\paragraph{Agents}\label{sec:bigraphs-agents}
Bigraphs with the domain $\epsilon$ are called \textit{agents} and are denoted with lower case letters of the form $a\colon \epsilon \to I$, often written $a\colon I$ (see \cite[p.~18]{milner_bigraphical_2001-1} and \cite[p.~74]{milner_space_2009}). Thus, agents are called \textit{ground}, meaning, they have no inner names and no sites. The initial set of bigraphs in a BRS are agents.

\paragraph{Reaction Rules} Reaction rules describe the semantics of a bigraphs. We distinguish between \textit{ground} and \textit{parametric} reaction rules.

\begin{definition}{(Parametric Reaction Rule (after \cite[p.~91]{milner_space_2009}))}\label{def:bigraph-reaction-rule}
	A parametric reaction rule for bigraphs is a triple of the form 
	\begin{equation*}
	(R\colon m \to J, R'\colon m' \to J, \eta\colon m' \to m)
	\end{equation*}
	where $R$ is the \textit{parametric redex}, $R'$ the \textit{parametric reactum}, and $\eta$ a map of finite ordinals. $R$ and $R'$ must be lean, and $R$ must have no idle roots or names. The rule generates all ground reaction rules $(r,r')$, where
	\begin{equation*}
	r \bumpeq R.d, \qquad r' \bumpeq R'.\bar{\eta}(d)
	\end{equation*}
	and $d\colon \langle{m,Y}\rangle$ is discrete.\footnote{A bigraph is called \textit{lean} if it has no idle edges \cite[p.~26]{milner_space_2009}. A bigraph is called \textit{discrete} if all its links are open and distinct (i.e., its link map is bijective). We write $f \bumpeq g$ for arrows in the same homeset if there is a bijection $\rho\colon |f| \to |g|$ that respects the structure of $f$ (see \cite[p.~23]{milner_space_2009}), where $|f|$ denotes the \textit{support} of a bigraph (see \cite[Def.~2.4]{milner_space_2009}).
	}
\end{definition}

The parameter $d$ is a discrete bigraph of the form $d = d_0 \otimes \cdots \otimes d_{m-1}$; $\eta$ the \textit{instantiation map} where $\bar{\eta}(d) = d_{\eta(0)} || \cdots || d_{\eta(m'-1)}$.

In order to evolve a bigraph, reaction rules must be applied to it. Considering the parametric reaction rule $\mathsf{R} = (R,R')$, $R$ is the search pattern to be found in an agent (i.e., the target bigraph). This problem is known as the \textit{bigraph matching problem}. %

\paragraph{Simple and Nice BRS}
The purpose of our presented approach is to generate agents for a certain class of BRSs. For this reason, we wish to give the definition of a particular class of BRSs termed \textit{simple and nice}. The imposed constraints are not severe and only produce a certain class that makes the LTS more tractable without losing expressiveness. For a detailed explanation with proofs, the reader may refer to \cite[p.~96]{milner_space_2009}.
This leaves us with the formal definition of a general BRS and its special class:

\begin{definition}{(Bigraphical Reactive System (BRS) (after Definition 8.6 \cite[p.~92]{milner_space_2009}))}\label{def:bigraph-brs}
A \textit{(concrete) bigraphical reactive system (BRS)} over $\Sigma$ consists of $`BG(\Sigma)$ equipped with a set $`\mathcal{R}$ of parametric reaction rules closed under support equivalence; that is, if $R \bumpeq S$ and $R' \bumpeq S'$ and $`\mathcal{R}$ contains $(R,R',\eta)$, then it also contains $(S,S',\eta)$. We denote the BRS by $`BG(\Sigma, `\mathcal{R})$. It is \textit{safe} if its sorting $\Sigma$ is safe.
\end{definition}

\begin{definition}{(Simple and Nice BRS (after Definition 8.12 and 8.18 \cite[pp.~95]{milner_space_2009}))}\label{def:bigraph-brs-nice}
A parametric redex is \textit{simple} if it is \textit{open} (every link is open), \textit{guarding} (no site has a root as parent) and \textit{inner-injective} (no two sites are siblings). 

A parametric redex is \textit{unary} if its outer face is. A reaction rule is \textit{simple}, or \textit{unary}, if its redex is so. A BRS is \textit{simple}, or \textit{unary}, if all its reaction rules are so.

A reaction rule is \textit{nice} if it is \textit{safe}, \textit{simple}, \textit{unary}, \textit{affine} and \textit{tight}. A BRS $`BG(\Sigma, `R)$ is \textit{nice} if all its reaction rules are nice.
\end{definition}

\section{Random Bigraph Generation}\label{sec:algorithm}

We come now to our algorithm for generating random bigraphs, which covers the construction of bigraph-compatible agents (recall the definition from \cref{sec:bigraphs-agents}).\footnote{The authors working on a framework, implemented with the Java programming language, for the creation and simulation of bigraphs. The proposed algorithm is also implemented therein.} 
Beginning with our algorithmic approach for place graph generation, we briefly detail the originating concept before introducing our link graph generation algorithm. In regards to the node connection, we cover two edge cases for the linking process of nodes for the link graph.

\subsection{Place Graph Generation}

\paragraph{Barabási-Albert-Model}

We adopt the Barabási-Albert (BA) model \cite{barabasi_emergence_1999} as the underlying framework for our purpose to generate random place graphs.
The model is widely used among researchers of large networks \cite{newman_structure_2003, clauset_power-law_2009}, at the same time simple in contrast to other graph models such as \cite{eppstein_steady_2002} or \cite{r._palmer_generating_2000}. 
Barabási et al. \cite{barabasi_emergence_1999} propose a random graph generation algorithm for scale-free networks whose degree distribution asymptotically obeying a power-law distribution.
This is also known as the preferential attachment model. Vertices in a graph that have more connections than others will form larger clusters than vertices with fewer edges. To clarify, the interconnection between vertices is not uniformly distributed. A new vertex has a higher probability of being connected to a vertex that already has a large number of connections (see \cite{barabasi_emergence_1999}).

We are recalling the algorithm from \cite{barabasi_emergence_1999} for completeness here. %
Let $G = (V,E)$ be an unlabeled network with a vertex set $V$, and an edge set $E$. At the first time step, the network is initialized with a small number of starting nodes $m_0$. Then at every subsequent time step, a new node is added and connected to $m$ different existing vertices (selection is subject to a certain probability) in $G$ by $m$ edges ($m \leq m_0$).
The connection of a new vertex $v$ to an existing vertex $v_i$ depends on the connectivity $d(v_i)$ (i.e., degree) of $v_i$. So the probability that $v$ is added to $v_i$ is $\mathrm{Prob}(v_i) = \dfrac{d(v_i)}{\sum_{j} d(v_j)} = \dfrac{d(v_i)}{|E|}$ (see \cite[p.~511]{barabasi_emergence_1999}). Effectively, this implements the preferential attachment feature because vertices with more "children" tend to get more connections than those with fewer "children". %

\paragraph{Algorithm}
We use a slightly different notion concerning place graphs when referring to "connections". Owing to the well-defined meaning of nesting of place graphs within the bigraph theory, we resist using the term "connection" here (though the concept is similar), and use the relation \textit{has-children}. We adopt the former algorithm and introduce several changes:
\begin{itemize}
	\item The number of trees of a place graph can be specified by the parameter $t$ indicating the initial nodes for $F^P$ (instead of $m_0$). Then, $n$ indicates the overall number of places (containing only roots and nodes in this case), and $m = n-t$ represents the number of nodes in $F^P$.
	\item We discard the part regarding the connection of a freshly created node to $m$ existing nodes at every time step. Here, we consider a tree where a child node has only one parent.
	\item We include \textit{control selection} when creating a node by randomly assigning some control  $C_{k} \in C$ which is drawn according to the discrete uniform distribution $k = \mathrm{unif}\{0, \vert C\vert-1\}$.%
	\item In the original algorithm, we can observe that for an isolated vertex $v_i$, the attachment probability is always $\mathrm{Prob}(v_i)=0$. We modify this property by giving each existing vertex a positive probability by incrementing the vertex degree by one, so that the attachment probability is always $\mathrm{Prob}(v_i) \geq 1$.\footnote{A similar approach was also proposed within the discontinued graph framework \textit{JUNG} \cite{omadadhain_analysis_nodate}, where the authors used a Lagrangian smoothing method to avoid a non-negative attachment probability.
	}
	However, we employ a "trick" to avoid the additional probability computation which determines if a new node shall be nested within another one by keeping an additional list of node references.\footnote{The idea is adopted from \cite{michail_jgrapht_2019}.
	}
	That means nodes appearing more frequently in the list corresponds to nodes having a higher degree. As a result, the likelihood of being selected as a parent increases.
\end{itemize}

\begin{algorithm}[h]
	\caption{Algorithm for generating random place graphs for a given signature.}
	\label{algo:placegraphgeneration}
	\begin{algorithmic}[1]
		\Procedure {PGG}{$t$, $n$, $C$}
		\State $places \gets \emptyset$\;
		\State $i \gets 0$\;
		\While{$|places| < t$}
			\State $places \gets places~\uplus~$ \textsc{Create\_Root}($i$)\;\label{algo:line:createNewRoot}
			\State $i \gets i + 1$\;
		\EndWhile
		\For{$i \gets t; i < n; i \gets i + 1$}
			\State $r \gets \mathrm{unif}\{0, |places| - 1\}$\;
			\State ${k} \gets \mathrm{unif}\{0, |C| - 1\}$\;
			\State $v \gets$ \textsc{Create\_Node}(${C}_{k}$)\;\label{algo:line:createNewNode}
			\State $prnt_F(v) \gets places_r$\;
			\State $places \gets places~\cup~v$ \;
			\If{$i > 1$}
				\State $places \gets places~\uplus~places_r$ \;
			\EndIf
		\EndFor
		\State \Return $F_P = (V_F, ctrl_F, prnt_F): \emptyset \to t$\;
		\EndProcedure
	\end{algorithmic}
\end{algorithm}

The procedure is shown in \cref{algo:placegraphgeneration} and shall be self-descriptive. However, we want to highlight some facts. The method \textsc{Create\_Root(index: int)} in \cref{algo:line:createNewRoot} creates a new root and takes an integer as argument indicating the index of this place. A control is randomly drawn from $C$ for the node creation. Therefore, the method \textsc{Create\_Node(c: control)} in \cref{algo:line:createNewNode} is responsible and accepts a control as argument. The procedure is executed until $n$ nodes in total are being created (including roots and nodes).

\subsection{Link Graph Generation}\label{sec:link-graph-generation}
The generation of a link graph depends on the support of a place graph (see \cite[Def.~2.4]{milner_space_2009}) and the signature. For a given place graph of a bigraph over the signature $\mathcal{K}$, we want to present how links between nodes may be added. Therefore, we offer two strategies to define constraints on linkings between nodes:
\begin{enumerate}[i)] %
	\item \textit{{Minimal Pairwise Port Linkage (Uniformly-Distributed)}}, and
	\item \textit{{Maximal Degree Correlation (Assortative or Disassortative)}}.
\end{enumerate}
For all strategies, we do not consider connections exclusively between inner names or allowing idle links and idle inner names in general. This is due to the fact that we focus on the agent construction of nice and simple BRSs (refer to \cref{sec:bigraphs-dynamic}).

\paragraph{Notation}
Let us define some parameters first. We have $p$ the probability that any link (whether open or closed) is created. Then, $p_o$ is the probability that an outer name is created, similar is $p_e$ the probability for edge creation. Both parameters are considered as weights (see \cref{algo:linkgraphgeneration}, \cref{algo:line:wr}).

\subsubsection{Minimal Pairwise Port Linkage}\label{sec:link-graph-generation:pairwise-port-linkage}

For our first strategy, we allow only \textit{one link connection} (i.e., through an outer name or an edge) between a pair of nodes. The connection is enabled using the node's port. It is obvious that the linking is not reflected upon roots since only nodes have ports. The procedure is presented in \cref{algo:linkgraphgeneration} and explained in the following.

\begin{algorithm}[h]
	\caption{Pseudo-code for generating a random link graph with minimal pairwise port linkage.}
	\label{algo:linkgraphgeneration}
	\begin{algorithmic}[1]
		\Procedure {MPPL}{$nodes$, $p$, $p_o$, $p_e$} \Comment{all nodes must have assigned a control with positive arity}

		\State $L\gets \lfloor{ 0.5 \, p \, \abs{nodes}}\rfloor$\;\label{algo:line:number-of-links}
		\If{$L < 1$}
			\State {\Return "probability $p$ is to small or to few nodes for creating links"}
		\EndIf
		\State $c \gets 0$\;
		\While{$c < L$}
		\State assert $i \neq j$: $i \gets \mathrm{unif}\{0, \abs{nodes} - 1\} \wedge j \gets \mathrm{unif}\{0, \abs{nodes} - 1\}$\;
		\If{\textsc{Weighted\_Random}($p_o$, $p_e$) $= 0$}\;\label{algo:line:wr}
			\State $l \gets $ \textsc{Create\_Outer()}\;\label{algo:line:createOuter}
			\State $Y \gets Y \uplus l$ \;
		\Else
			\State $l \gets $ \textsc{Create\_Edge()}\;\label{algo:line:createEdge}
			\State $E_F \gets E_F \uplus l$
		\EndIf
		\State $link_F((nodes_i, 0)) \gets l$\;\label{algo:line:linkNodes}
		\State $link_F((nodes_j, 0)) \gets l$\;\label{algo:line:linkNodes2}
		\State $nodes \gets nodes - \{nodes_i, nodes_j\}$\;
		\State $c \gets c + 1$\;
		\EndWhile\;
		\State \Return $F^L = (nodes, E_F, ctrl_F, link_F): \emptyset \to Y$\;
		
		\EndProcedure
	\end{algorithmic}
\end{algorithm}

The first step is to determine the maximal number of possible pairwise links $L$ (\cref{algo:linkgraphgeneration}, \cref{algo:line:number-of-links}) that can be created between the nodes. The fraction of the number of links to create, can be tuned by adjusting the parameter $p \in [0,1]$. The function \textsc{Weighted\_Random(a: double, b: double)} (\cref{algo:line:wr}) samples a value from the tuple $(0,1)$ with the given probabilities \textsc{a} and \textsc{b} corresponding to the value $0$ and $1$ of that tuple, respectively. For example, let $a=0.3$ and $b=0.8$ then there is a probability of $30\,\%$ to create an outer name (0) and a probability of $80\,\%$ to create an edge (1). The computed value of that function is evaluated in the \verb|if-then| construct to determine which link shall be created. The link $l$ in question is either an outer name returned by the method \textsc{Create\_Outer()} (\cref{algo:line:createOuter}), or an edge created by the method \textsc{Create\_Edge()} (\cref{algo:line:createEdge}). Both methods assign a unique label to that link. After, both variables $nodes_i$ and $nodes_j$ are being linked to $l$ (\cref{algo:line:linkNodes} and \cref{algo:line:linkNodes2}). For convenience, we use the first port of each selected node (refer to \cref{def:linkGraph}). The linked nodes are then removed from $nodes$ and ensures that no node is linked multiple times. This procedure is repeated until no more pairwise linkings are possible.

\subsubsection{Maximal Degree Correlation}\label{sec:link-graph-generation:maximal-degree-correlation}
The second model aims to achieve the highest possible linkage between nodes under a given bigraph's signature. To obtain another perspective on this task, we regard the given controls with their assigned arities as \textit{degree sequence} from which we want to create a link graph. Therefore, several models exist such as the \textit{hidden parameter model} or the \textit{configuration model} (see \cite{barabasi_network_2016}).
Furthermore, we want to influence the link graph's \textit{assortative} or \textit{disassortative} feature. Assortativity is a property indicating the preference that nodes are more likely to connect to similar nodes.
We must ensure that no self-loops or idle links are generated. We cannot create infinite links between nodes as the arity of a node is fixed and naturally functions as a constraint.

\paragraph{Xulvi-Brunet and Sokolov model}
Our following model is influenced by the \textit{Xulvi-Brunet and Sokolov} model \cite{xulvi-brunet_reshuffling_2004, xulvi-brunet_changing_2005}. It allows adjusting the magnitude of degree correlations for being either more assortative or disassortative. Therefore, their model employs a parameter $p \in [0,1]$ to vary the network's magnitude of being fully assortative ($p=1$) or fully random ($p=0$). Consequently, it always allows us to generate maximal degree correlations for both "directions".

\paragraph{Algorithm}\label{sec:algo:linkgraphgeneration-2}

Analogously in terms of link graphs, assortativity means that nodes with many ports tend to connect to similar nodes with high port numbers as well as nodes with small arity tend to connect to nodes whose controls also exhibit small arities. In contrast to disassortative mixing, nodes with high arity tend to connect to nodes with lower arity and vice versa.

In the parlance of \cite{barabasi_network_2016,xulvi-brunet_reshuffling_2004,xulvi-brunet_changing_2005}, we now explain the individual steps of our adapted algorithm for link graphs at this point instead of printing pseudo-code.
Given $F^L$ with $V_F$ where each node is assigned a control with positive arity, all nodes are being put in a working queue $Q$ at the beginning. Then, the following steps are executed: 
\begin{enumerate}[Step 1.]
	\item Randomly select $4$ different nodes where each one has at least one free port.
	\item We then apply a simplified wiring step. Based on the desired degree correlation, the \textit{wiring} is \textit{assortative or disassortative}. For the former, we link both the nodes with the highest degree, followed by the remaining two nodes with the lowest degree. For the latter, we link the node with the highest and lowest degree, and finally, the two remaining ones. For linkage, a new edge is created, and two nodes are connected to it via their free port. That means an edge will have at most two vertices connected.
	\item If a node has no free ports, it is removed from $Q$, thus, not available anymore for linkage. The algorithm is repeated, starting from Step 1, as long as the number of elements in $Q$ is greater or equal than $4$. 
\end{enumerate}

Opposed to \cite{xulvi-brunet_reshuffling_2004,xulvi-brunet_changing_2005}, we are removing nodes from $Q$, which is a natural constraint within the framework of bigraphs as the arity specifies the maximum number of links a node can have. In this case, the algorithm terminates automatically if $Q$ has less than four nodes.

\section{Experimentation and Analysis}\label{sec:analysis}

\subsection{Place Graph Analysis}\label{sec:analysis-placegraphs}
The structure of a place graph can exhibit complicated patterns realized by their parent-child-relationship.
Here, we study several properties of some artificially created place graphs, including the degree distribution and the distribution regarding the arity of the nodes.
The following conducted place graph analyses refer exclusively to \Cref{algo:placegraphgeneration}.

\subsubsection{On the Degree Distribution}\label{sec:On-the-Degree-Distribution}

The number of children and parents of a node in a place graph $F^P$ is called the degree of a node $d(v) = \abs{prnt_F(v)} + \abs{prnt_F^{-1}(v)}$ where $prnt_F^{-1}(v)$ returns the children of $v$. Thus, the degree of the node denotes the size of the \textit{open neighborhood} of a node. With respect to pure bigraphs, a node can only have one parent, so $\abs{prnt_F(v)} = 1$.

The first insight we obtain into $F^P$ is by computing the frequency of distinct node degrees. Let $\text{deg}(d)$ be the corresponding degree distribution:
\begin{gather*}
\text{deg}(d) \eqdef \text{fraction of nodes in the place graph with degree $d$}.
\end{gather*}

\begin{figure}[htbp]
	\centering
	\includegraphics[width=0.85\textwidth]{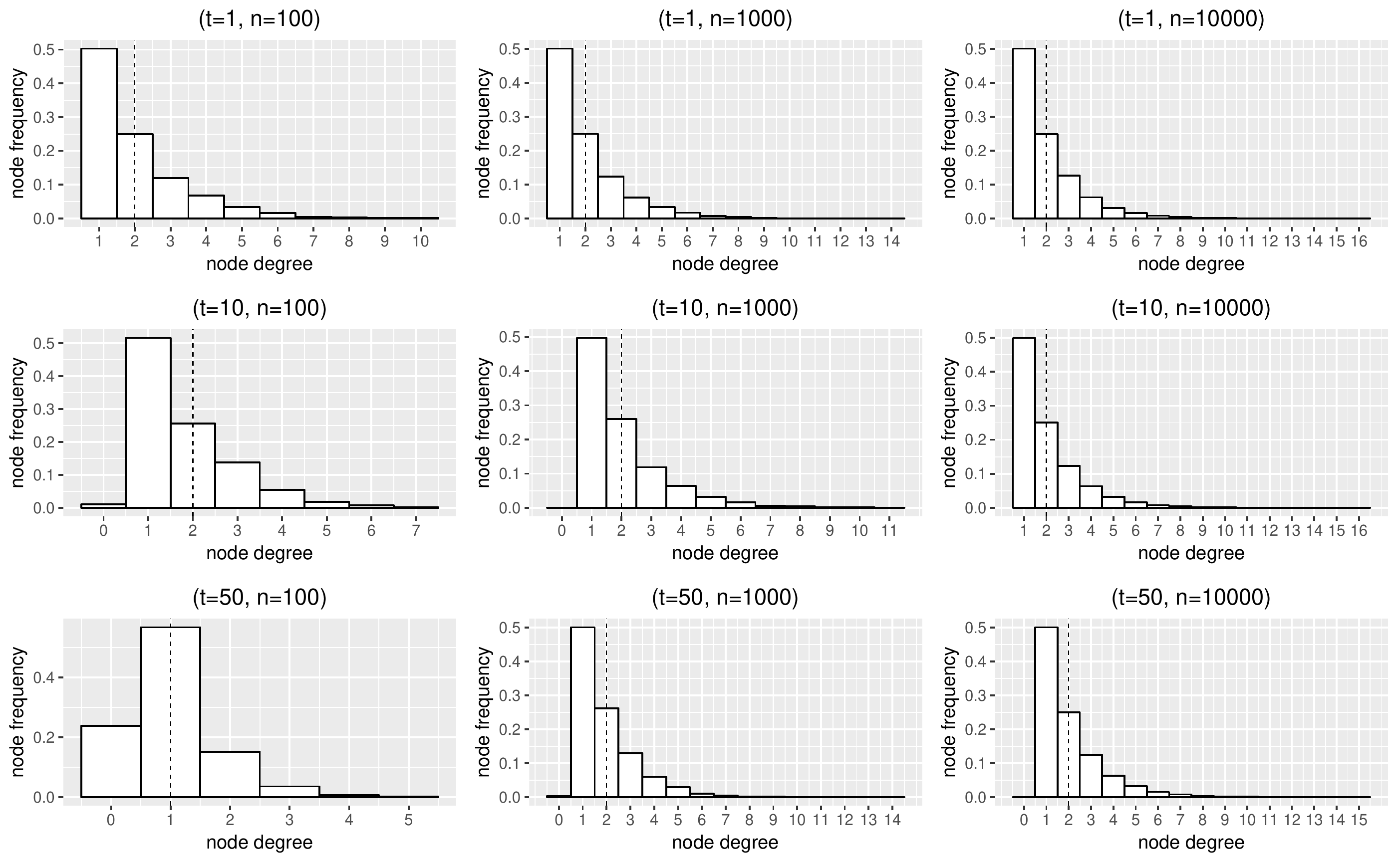}
	\caption{Histograms of the node degree distribution for several place graphs with a varying number of roots $t$ and nodes $n$ are shown. A dashed vertical line marks the average node degree for each configuration.}
	\label{fig:analysis-placegraph-degree-distribution}
\end{figure}

To assess the average degree distribution, we sample some place graphs. Therefore, we chose $t=\{1, 10, 50\}$ for the number of trees and $n = \{10, 100, 1000\}$ for the number of nodes. For each parameter combination $c_{t_i,n_j} \in t \times n$ we performed $r=100$ independent runs to generate a sufficiently large sample population. The results are averaged and presented in \Cref{fig:analysis-placegraph-degree-distribution}.

From \cref{fig:analysis-placegraph-degree-distribution} we can observe that only very few nodes have many children (to the right of the histogram is the long tail, the node frequency is small) whereas the greater number of nodes have less children itself (to the left of the histogram are the nodes that dominate, the node frequency is high). Such distributions are identified by their long tails, which mainly reflect the \textit{preferential attachment property}.

The average degree across all configurations is relatively constant, with a value of approximately $2$. In other words, node nesting is not performed independently as it is analogously the case for the ER model, where edges between nodes are created independently. However, an exception is noted for $c_{50, 100} = 1$.
This is not surprising because our algorithm always creates roots first, and consequently, half of the places are already root nodes, more or less "forcing" the remaining nodes to be nested under one of these roots. The first chosen ones in the early phase quickly turn the remaining roots to orphans, thus, becoming increasingly less likely to be selected. Many of them have a node degree of zero (i.e., no children), which explains the left-shift of the average degree compared to the other configurations.

\subsubsection{On the Node's Arity Distribution}

Now, we analyze the node distribution by taking the node's port count into consideration.
Recalling \cref{algo:placegraphgeneration}, we have chosen a uniform distribution for control selection, ensuring a highly diverse place graph w.r.t. to a node's control. Of course, one can choose a different distribution and assign to each control a different probability of occurrence. In any case, the control selection process is independent of their respective labels and arity. This kind of analysis we conduct shortly is especially essential, as it impacts the execution of the link generation model (refer to \cref{sec:link-graph-generation} and \cref{sec:link-graph-analysis}). The interconnection between nodes is only viable if their port count is not zero.

Therefore, we ask the following question: How high is the probability that $k$ nodes with positive arity are in some generated $F^P$ over the signature $\Sigma$ with $n$ nodes, when the fraction of controls with positive arity in $\Sigma$ is $p \in [0,1]$? 

\begin{claim}{}\label{def:claim0}
	A concrete place graph $F^P$ over $\Sigma$ is randomly generated with $n$ nodes (precluding roots and sites).
	Let us define $p$, the probability of controls with $ar(C_i) \geq 1$, and $q = 1 - {p}$ for controls with $ar(C_i) = 0$, where $C \in \Sigma$. This separates $C$, regardless of its labels, into two groups.
	We then have for $\vec{v} = \{v_i\,|\,ar(ctrl(v_i)) \geq 1\}$ that ${p} = \frac{\abs{\vec{v}}}{n}$.
	Denote by $X$ the stochastic variable for the number of nodes with positive arity.
	Then $X$ obeys a binomial distribution $X \sim B_{n,p}(k)$ with probability function ${n \choose k}\,p^k\,(1-p)^{n-k}$. 
	This means that the expected number of nodes having positive arity is $np$ with mean variance $np(1-p)$.

\end{claim}

\begin{figure}
	\centering
	\begin{subfigure}[b]{0.825\textwidth}
		\centering
		\includegraphics[width=\textwidth]{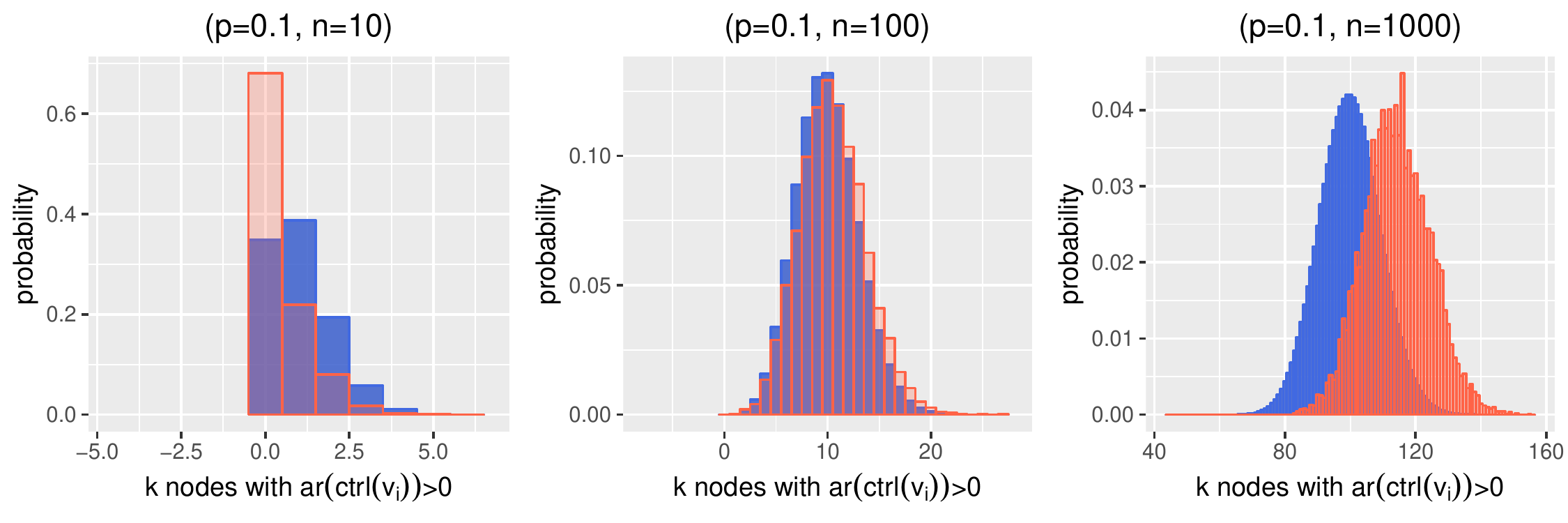}
		\caption{Barplots of both density distributions for $p=0.1$ and each $n_i$.}
		\label{fig:y equals x}
	\end{subfigure}
	\vfill
	\begin{subfigure}[b]{0.825\textwidth}
		\centering
		\includegraphics[width=\textwidth]{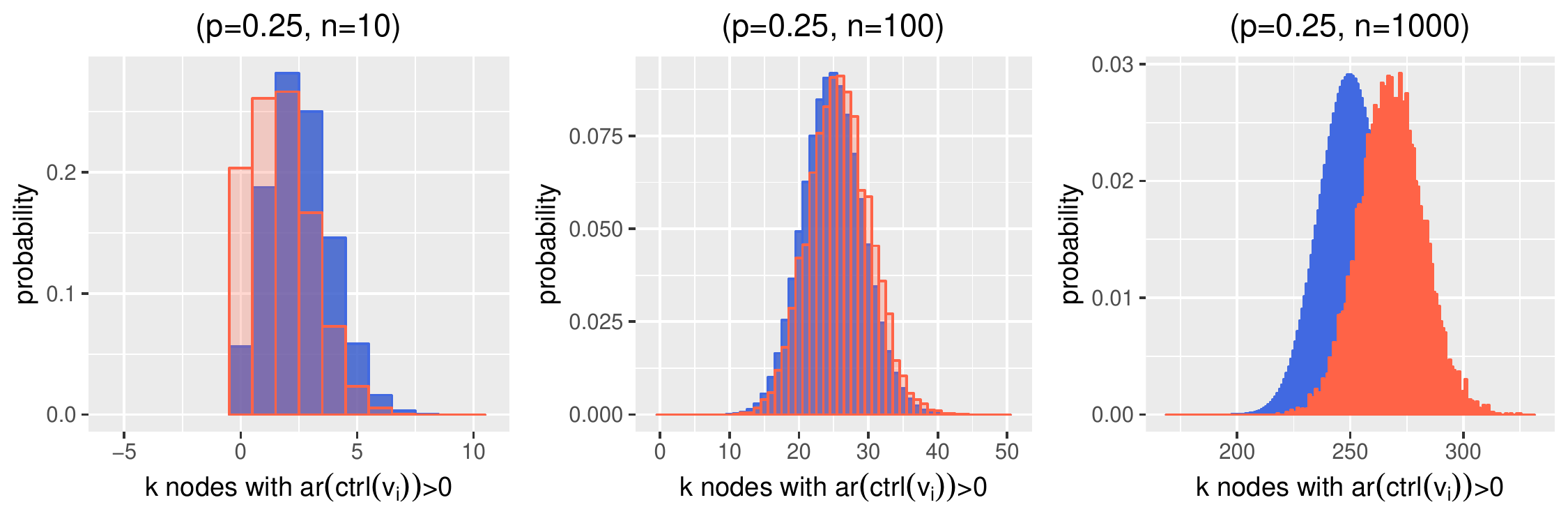}
		\caption{Barplots of both density distributions for $p=0.25$ and each $n_i$.}
		\label{fig:three sin x}
	\end{subfigure}
	\vfill
	\begin{subfigure}[b]{0.825\textwidth}
		\centering
		\includegraphics[width=\textwidth]{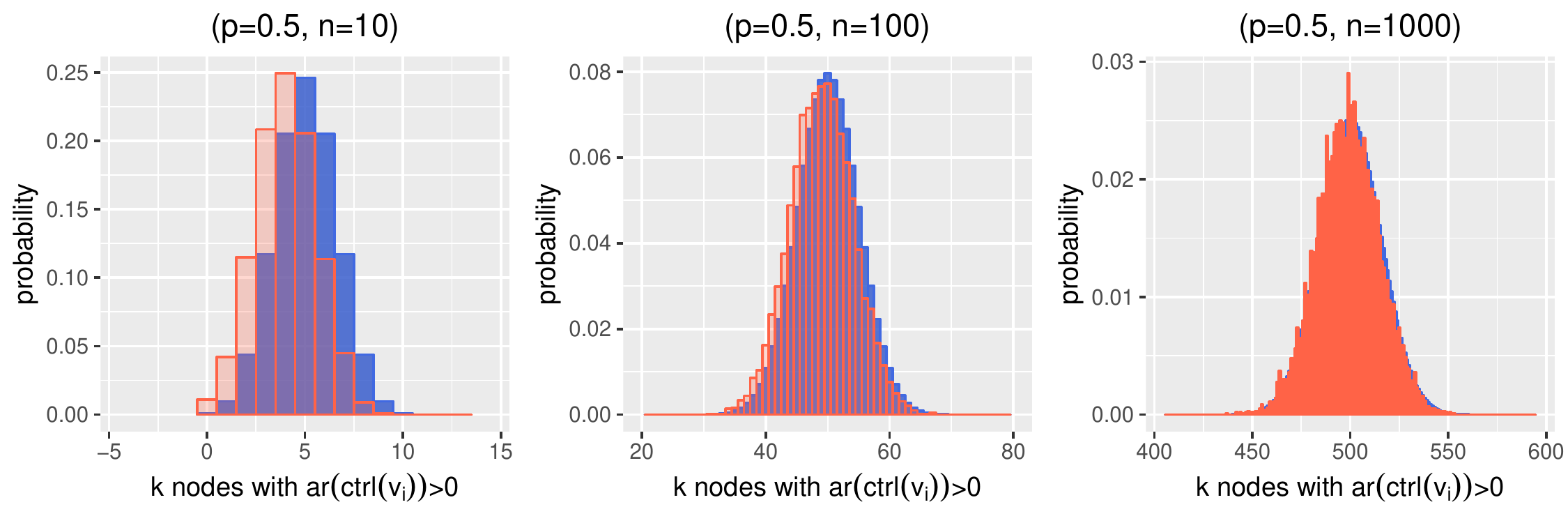}
		\caption{Barplots of both density distributions for $p=0.5$ and each $n_i$.}
		\label{fig:five over x}
	\end{subfigure}
	\vfill
	\begin{subfigure}[b]{0.825\textwidth}
		\centering
		\includegraphics[width=\textwidth]{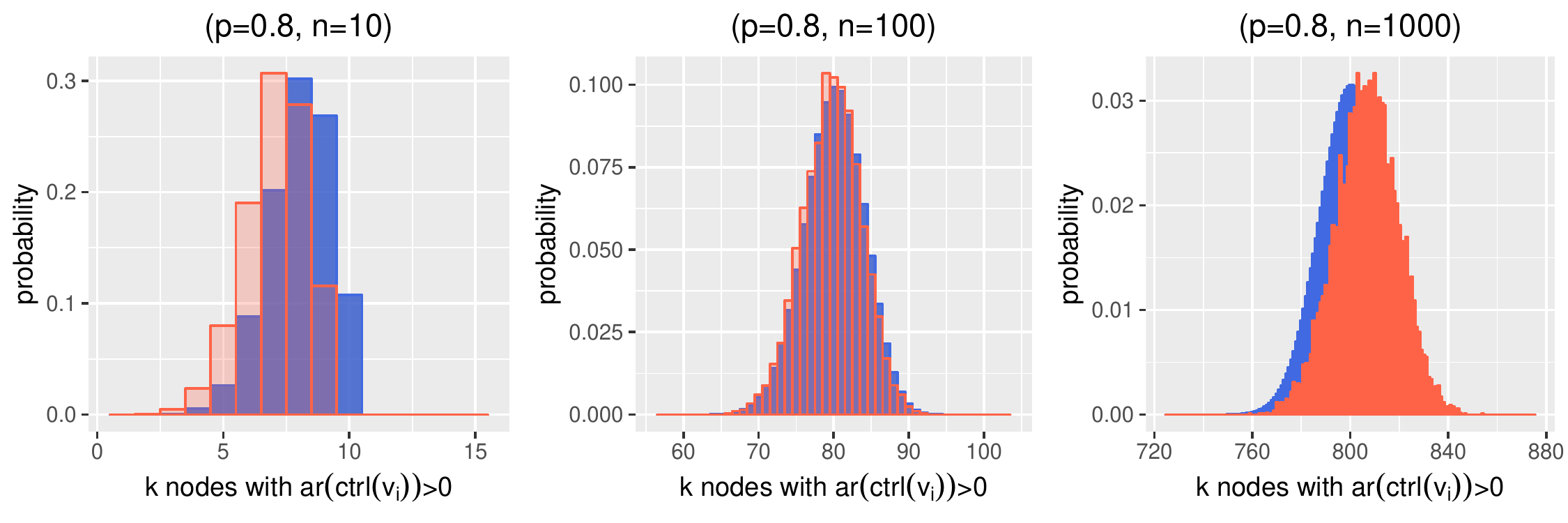}
		\caption{Barplots of both density distributions for $p=0.8$ and each $n_i$.}
		\label{fig:five over x2}
	\end{subfigure}
	\caption{Barplots of the theoretical (blue) and experimental (orange) distribution for each parameter configuration $c_{p_i,n_j}$. }
	\label{fig:analysis-placegraph-distribution}
\end{figure}

To prove our claim, we compare the theoretical quantiles of the binomial distribution against our experimental data.
Therefore, we vary 
\begin{inparaenum}[i)]
	\item the number of nodes, and,
	\item the fraction of controls with a positive arity from a set of $26$ controls. 
\end{inparaenum}
We fix $t=1$ (number of roots) and vary the number of nodes $n = \{10, 100, 1000\}$ and the fraction $p = \{0.1, 0.25, 0.5, 0.8\}$ of controls assigned a positive arity from a signature. For each parameter combination $c_{p_i, n_j} \in p \times n$, we performed $r=10000$ independent runs for generating a sufficiently large sample population, which we averaged at the end. The results are summarized in \cref{tab:analysis_placegraph}; the density distribution of the experimental data and the theoretical binomial distribution are visualized in \cref{fig:analysis-placegraph-distribution} for all parameter configurations $c_{p_i, n_j}$.

\begin{table}
	\centering
	\caption{Experimental values in comparison with the statistical measures of the binomial distribution. The corresponding theoretical statistical measures are within parenthesis next to the computed ones. The expected value $E[X]$ is $np$, the mean variance $\mathrm{Var}[X]=np(1-p)$.}\label{tab:analysis_placegraph}
	\begin{tabular*}{\textwidth}{@{\extracolsep{\fill}} l l l l l l l}
		\toprule
		Probability & $\abs{V_F}$ & $\mathrm{E}[X]$ & $\mathrm{Sd}[X]$ & $\mathrm{Var}[X]$ & Skewness & Kurtosis \\ 
		\midrule
		\multirow{3}{*}{$p_1=0.1$} & $n_1=10$ & 1.14 (1) & 1.02 (0.95) & 1.03 (0.9) & 0.77 (0.84) & 0.35 (0.51) \\ 
		& $n_2=100$ & 11.52 (10) & 3.19 (3) & 10.16 (9) & 0.28 (0.27) & 0.13 (0.05) \\ 
		& $n_3=1000$ & 115.3 (100) & 10.08 (9.49) & 101.53 (90) & 0.12 (0.08) & -0.04 (0.01) \\
		\midrule 
		\multirow{3}{*}{$p_2=0.25$} & $n_1=10$ & 2.7 (2.5) & 1.41 (1.37) & 1.98 (1.88) & 0.33 (0.37) & -0.06 (-0.07) \\ 
		& $n_2=100$ & 26.87 (25) & 4.4 (4.33) & 19.4 (18.75) & 0.11 (0.12) & -0.02 (-0.01) \\ 
		& $n_3=1000$ & 269.17 (250) & 13.97 (13.69) & 195.22 (187.5) & 0.05 (0.04) & 0.05 (0) \\ 
		\midrule
		\multirow{3}{*}{$p_3=0.5$} & $n_1=10$ & 5 (5) & 1.57 (1.58) & 2.47 (2.5) & 0 (0) & -0.16 (-0.2) \\ 
		& $n_2=100$ & 50.02 (50) & 5.07 (5) & 25.75 (25) & 0.01 (0) & -0.06 (-0.02) \\ 
		& $n_3=1000$ & 499.91 (500) & 15.69 (15.81) & 246.31 (250) & 0.01 (0) & 0.01 (0) \\ 
		\midrule
		\multirow{3}{*}{$p_4=0.8$} & $n_1=10$ & 8.07 (8) & 1.25 (1.26) & 1.56 (1.6) & -0.5 (-0.47) & 0.09 (0.03) \\ 
		& $n_2=100$ & 80.73 (80) & 3.91 (4) & 15.3 (16) & -0.15 (-0.15) & -0.03 (0) \\ 
		& $n_3=1000$ & 807.83 (800) & 12.63 (12.65) & 159.41 (160) & -0.07 (-0.05) & -0.05 (0) \\ 
		\bottomrule
	\end{tabular*}
\end{table}

\begin{table}[htbp]
	\centering
	\caption{The table shows the parameter estimation by MLE and their goodness-of-fit by AIC. The binomial, Poisson, and geometric discrete probability distributions were tested. The estimated parameter $\hat{\theta}$ for each distribution is shown with its standard error $\mathrm{SE}$ and the corresponding result of the negative log-likelihood function $\mathcal{L}(\hat{\theta})$. The statistical measure AIC (with $k=1$) for the corresponding $\mathcal{L}(\hat{\theta})$ is shown. For the Poisson distribution we use the estimate $\mu = np_i$. The two estimates $p_3=0.5,p_4=0.8$ for the geometric distribution could not be computed by \texttt{fitdistr} and were computed manually.}\label{tab:analysis-placegraph-aic}
	\begin{tabular*}{\textwidth}{@{\extracolsep{\fill}} llllll}
		\toprule
		Model	& Model Parameters & Estimation $\hat{\theta}$ &  SE & $L(\hat{\theta})$ &  AIC \\ 
		\midrule 
		\multirow{4}{*}{Binomial Distribution} & $p_1=0.1, n=1000$ & 0.1153016 & 0.0001009928 & -37280.12 &  74562.24 \\ 
		& $p_2=0.25, n=1000$ & 0.2691707 & 0.0001402548 & -40558.07 & 81118.14 \\ 
		& $p_3=0.5, n=1000$ &  0.499906 & 0.0001581132 & -41722.31  & 83446.63\\ 
		& $p_4=0.8, n=1000$ & 0.8078311 & 0.0001245929 &  -39543.89  & 79089.77\\
		\midrule
		\multirow{4}{*}{Poisson distribution} & $\mu_1=100$ & 115.2991 & 0.1073774 & -37318.83 & 74639.65 \\   
		& $\mu_2=250$ & 269.17  & 0.164064 & -40790.42 & 81582.85 \\ 
		& $\mu_3=500$ & 499.906 & 0.2235858 & -42724.98 & 85451.97 \\ 
		& $\mu_4=800$ & 807.8324 & 0.2842239 &-43648.64 & 87299.28 \\ 
		\midrule
		\multirow{4}{*}{Geometric Distribution} & $p_1=0.1$ & 0.008598519 & 8.444783e-05 & -57518.54 & 115039.1 \\ 
		& $p_2=0.25$ & 0.003701373 & 3.398385e-05 & -65971.98 & 131946 \\ 
		& $p_3=0.5$ &  0.002 & 0.0001935838 & -72154.21  & 144312.4\\ 
		& $p_4=0.8$ & 0.0012 & 0.0001208009 & -76954.15  & 153912.3\\
		\bottomrule
	\end{tabular*} 
\end{table}

In addition to the results from \cref{tab:analysis_placegraph} and \cref{fig:analysis-placegraph-distribution}, we wish to conduct a test which distribution, among three candidate models, namely, \textit{binomial}, \textit{Poisson} and \textit{geometric} distribution, gives us the best fit for our experimental data for each configuration $c_{p_i, n_j}$. To conduct the test, we use the R package \verb|fitdistr| \cite{delignette-muller_fitdistrplus_2015}. We wish to briefly explain the procedure here to avoid common misconception.\footnote{Remark: The employed method does not tell us how well it fits our data, instead it is used to compare the best fit among a set of distribution models (see also \cite[p.~62]{burnham_model_2002}).} The fit of a distribution is performed using maximum likelihood estimation (MLE). For an univariate parametric probability density function $f({x}_i|\mathbf{\theta})$, the aim is to find the best estimate for unknown $\theta$. The best estimate $\hat{\theta}$ is the value which maximizes the likelihood function $\mathcal{L}(\theta) = \prod_{i=1}^{n} {f(x_i|\theta)} = f(x_1|\theta) \cdot f(x_2|\theta) \cdots f(x_n|\theta)$. To find the optimum, we minimize the negative log-likelihood function in order to estimate
$
\hat{\theta} = \arg min_{\theta}~{- \ln{\mathcal{L}(\theta)}} = argmin_{\theta}~{- \sum_{i=1}^{n} {\ln{f(x_i|\theta)}}}
$.
We judge the model's goodness-of-fit by means of the \textit{Akaike information criteria} (AIC) for selecting a model candidate from our set of models which is computed as the maximum of $\mathcal{L}(\theta)$: $2k +  \ln{(\mathcal{L}(\theta))} - 2$ where $k$ is the number of estimable parameters (cf. \cite[p.~61]{burnham_model_2002}). The lowest value indicates the best model among the candidate models being used for evaluation. 
For reasons of clarity concerning the results, the AIC is computed only for $c_{p_i,n}, n=1000$ from the same experimental data set as shown in \Cref{tab:analysis_placegraph}. The results are presented in \cref{tab:analysis-placegraph-aic}.
Based on the AIC, we can opt for the binomial distribution among our three candidate distributions as it best fits the data. The larger $p_i$ is, the higher the difference of the AIC between the binomial distribution and each of the other two models becomes. 
This concludes the place graph analyses and leaves us with the link graph analyses.

\subsection{Link Graph Analysis}\label{sec:link-graph-analysis}

\subsubsection{Observations about the Minimal Pairwise Port Linkage Algorithm}\label{sec:analysis-pairwise-linkage}

In this section, the following statements about the link graph generation algorithm refer exclusively to \Cref{algo:linkgraphgeneration}. Therefore, we wish to study some properties of link graphs now.

\begin{lemma}{}\label{def:prop}
	The proportion of linked nodes to non-linked nodes of a link graph (with $n$ places in total, where $m$ nodes of $n$ have a positive arity) is 
	$$\frac{{p}\,m}{n} \propto \frac{n - {p}\,m}{n}$$
	with $p \in [0,1]$ being the fraction of nodes being selected for the pairwise linking among the nodes assigned a positive arity.
\end{lemma}

It is easy to verify that \Cref{def:prop} is true with regards to \Cref{algo:linkgraphgeneration}.

Now, we determine the total number of linked nodes the algorithm creates. If ${p}=1$, it is easy to see, that at least two nodes must exist to create a pairwise connection, both having some control $C_i$ with $ar(C_i) \geq 1$.
We can conclude from \Cref{algo:linkgraphgeneration} (refer to \Cref{algo:line:number-of-links}) that the \textit{number of pairwise links} $L$ among nodes $\{v_i\}_{i \in \{1,\ldots,m\}}$ of a link graph $F^L$ is:
\begin{equation}
\begin{gathered}
L = \Bigg\lfloor{\frac{pm}{2}}\Bigg\rfloor. \\
\end{gathered}
\end{equation}
To ensure \textit{pairwise linkage} ($L > 0$), the lower bound for the parameter $p$ is ${p} \geq {\frac{2}{m}}$ with $m \geq 2$.

\begin{proposition}{}\label{def:prop2}
	Let $X$ be the discrete stochastic variable describing the selection of a node for linkage and $f\colon i \to \mathbb{N}$ a mapping assigning every node index the number of ports of the corresponding node. We consider $m$ nodes subject to the condition $f(i) \geq 1, v_i \in V_F, i \in \{1,\ldots,n\}$.
	Then the probability $pr_k=Prob(X=k)$ for a node $v_k$ to be selected for linkage is $\frac{1}{m}$. This means that $Prob(X\geq 1)=1$.
\end{proposition}

Regarding \Cref{algo:linkgraphgeneration}, already connected nodes are removed from the set of available nodes, effectively decrementing $m$ by $2$ after every iteration. However, every node has the same probability of being selected for linking, which leaves the selection process independent. It is guaranteed that at least one node pair is found for linking when the lower bound is respected.

\subsubsection{On Assortativity of the Maximal Degree Correlation Algorithm}

The following analysis refers to the algorithm explained in \Cref{sec:algo:linkgraphgeneration-2} and investigates the question of whether there is any correlation between the node's ports and the assortativity of nodes.
Assortativity is a key measure for investigating the structure of a network \cite{fisher_perceived_2017}. Link graphs provide similar semantics compared to networks because their hyperedges allow the connection of multiple nodes to one edge. Therefore, the assortativity (also termed degree correlation) among all nodes is a measure indicating how the linkage behavior depends on the node's port count, and thus, the control's arity (cf. \cite{barabasi_network_2016,fisher_perceived_2017,thedchanamoorthy_node_2014}).
For such interpretation of the behavior of the link graph generation model presented in \cref{sec:link-graph-generation:maximal-degree-correlation}, we use an alternative measure from \cite{thedchanamoorthy_node_2014} to inspect the node's assortativity of a link graph, instead of using the \textit{degree correlation function} or the \textit{degree correlation coefficient} \cite{newman_assortative_2002} (for a detailed explanation of these two measures see \cite{barabasi_network_2016,newman_assortative_2002}). 
The reason is that the degree correlation coefficient may misestimate the assortativity in large networks \cite{fisher_perceived_2017}.

The alternative metric of assortativity of a network topology is rather seen as a relative concept by Thedchanamoorthy et al. \cite{thedchanamoorthy_node_2014}, which we wish to adopt here. We agree that it provides a more relevant perspective on the node assortativity distribution as it is often not meaningful considering networks either completely as assortative or disassortative \cite{thedchanamoorthy_node_2014}. We study the link graph's topology by classifying the assortativity of a link graph on the node level instead on the graph level because individual nodes can exhibit assortative and disassortative tendencies \cite{thedchanamoorthy_node_2014}. The indicator of the node's disassortativity (i.e., being non-assortative) for a link graph $F^L$ reads:
\begin{equation}\label{eq:average-neighbor-difference}
\delta_v = \frac{1}{d_v} \sum_{d_i\in neighbors_{L}(v)}^{}{\abs{d_i - d_v}}
\end{equation}
which is called the \textit{average neighbor difference} (cf. \cite[p.~2452]{thedchanamoorthy_node_2014}), where $d_v = \abs{link_F(ports(v))}$ with $ports(v) = \{(v, i) \,|\, i \in ar(ctrl_F(v))\}$ and $neighbors_{L}(v)= {link_F^{-1}(link_F(ports(v))) - ports(v)}$ for a link graph $F^L$ (refer to \Cref{def:linkGraph}). However, we omit the associated port of a node in each tuple $\in neighbors_{L}$ for \Cref{eq:average-neighbor-difference}; for example we write $neighbors_{L}(v) = \{v_{i}, \ldots, v_{i+n}\}$ for 
$\{(v_{i},0), (v_{i}, 1), \ldots, (v_{i+n}, 0)\}$.
It measures the number of differences concerning the degree of some node between the node's neighbors in terms of the link graph.\footnote{The node degree definition refers exclusively to link graphs here but is the same concept as introduced in \Cref{sec:On-the-Degree-Distribution} for place graphs.} Therefore, we take into account the number of actually connected ports instead of only using the arity of the node's control.

\begin{figure}[htbp]
	\centering
	\includegraphics[width=0.25\linewidth]{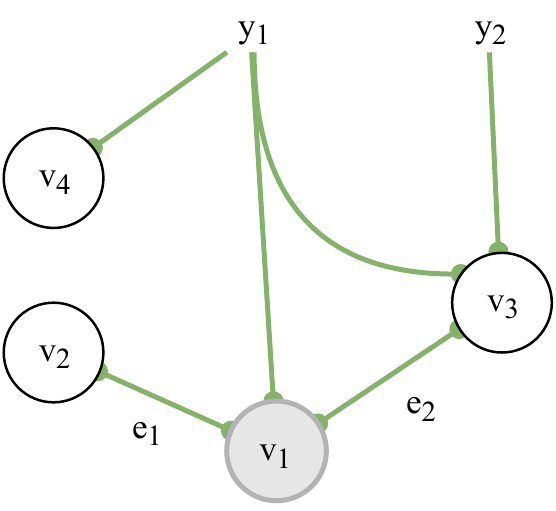}
	\caption{An arbitrary link graph with four nodes and two outer names. The average neighbor difference is a quantity which reflects the disassortativity of a node. For the given link graph, the average difference for the highlighted node $v_1$ is $\delta_{v_1} = 1.\overline{3}$.}
	\label{fig:link-graph-disassortativity-indicator}
\end{figure}

As an example, consider \Cref{fig:link-graph-disassortativity-indicator} which shows a link graph with four nodes and two outer names. Node $v_1$ has 3 ports which are connected to the edges $e1, e2$ and the outer name $y_1$, thus $d_{v_1} = 3$. Further it is linked to all remaining nodes, namely to $v_2$ by edge $e_1$, also to $v_3$ via edge $e_2$ and outer name $y_1$, and lastly to $v_4$ by $y_1$, thus $neighbors_{L}(v_1) = \{v_2, v_3, v_4\}$. The neighbors themselves have the degrees $d_{v_2} = 1, d_{v_3} = 3, d_{v_4} = 1$, hence, $\delta_{v_1} = {(\abs{1-3} + \abs{3-3} + \abs{1-3})} / {3} = 1.\overline{3}$.

Then, the values are scaled by dividing each $\delta_v$ by $S = \sum \delta_v$ so that the sum of the scaled values $\hat{\delta}_v$ is $S' = 1$. A scaling factor $\lambda = \frac{S'+r}{N} = \frac{1+r}{N}$ is introduced such that $\sum(\lambda - \hat{\delta}_v) = N\,\lambda - S' = r$ (see \cite[p.~2453]{thedchanamoorthy_node_2014}), where $N$ is the number of nodes and $r$\footnote{With $-1 \leq r \leq 1$, where $r=1$ indicates perfect assortativity and $r=-1$ indicates perfect disassortativity.} the correlation coefficient (see \cite{newman_mixing_2003}) to match the network's assortativity. Finally, the \textit{assortativity of a node} is calculated as
\begin{equation}\label{eq:alternative-assortativity}
\alpha_v = \lambda - \hat{\delta}_v.
\end{equation}
By definition, all assortative nodes yield $\alpha_v \geq 0$. We wish to give some explanatory remarks on the interpretation of \Cref{eq:alternative-assortativity} for the analysis we shortly conduct. We consider nodes with an assortativity within the interval of the deviation $+3\,\sigma_\alpha$ ($-3\,\sigma_\alpha$) as slightly more assortative (slightly more disassortative) than the distribution's average $\mu_\alpha$ in the link graph. As apparent from the above discussion, the metric explains the relative assortativity on the \textit{node level} instead of making assumptions on the higher \textit{network level}. The measure is defined for undirected networks which bigraphs are: the links of the link graph make no distinction between indegree and outdegree, and self-loops are not permitted.

\begin{figure}
	\begin{subfigure}[b]{0.49\textwidth}
		\centering
		\includegraphics[width=0.94\linewidth]{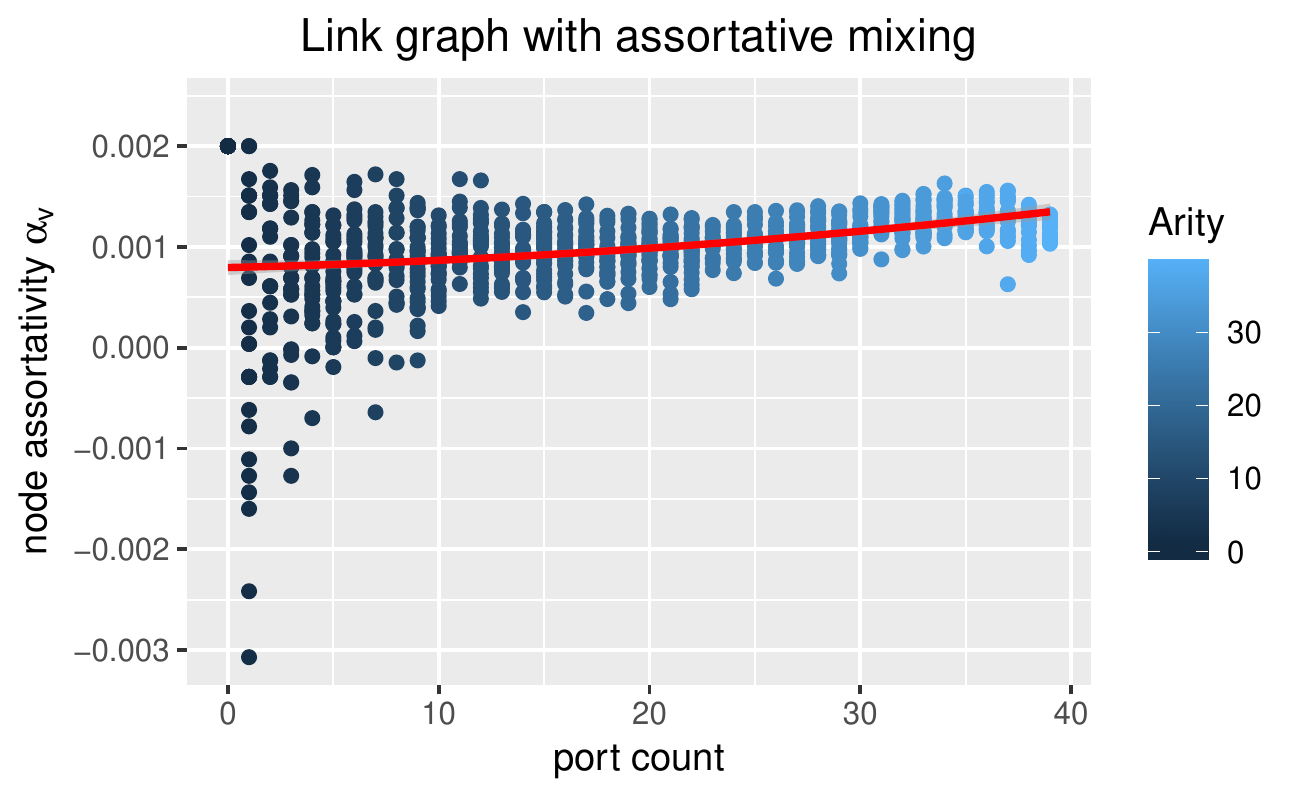}
		\caption{Node assortativity for a link graph with assortative mixing.}
		\label{fig:node-assortativity-a}
	\end{subfigure}
	\hfill
	\begin{subfigure}[b]{0.49\textwidth}
		\centering
		\includegraphics[width=0.94\linewidth]{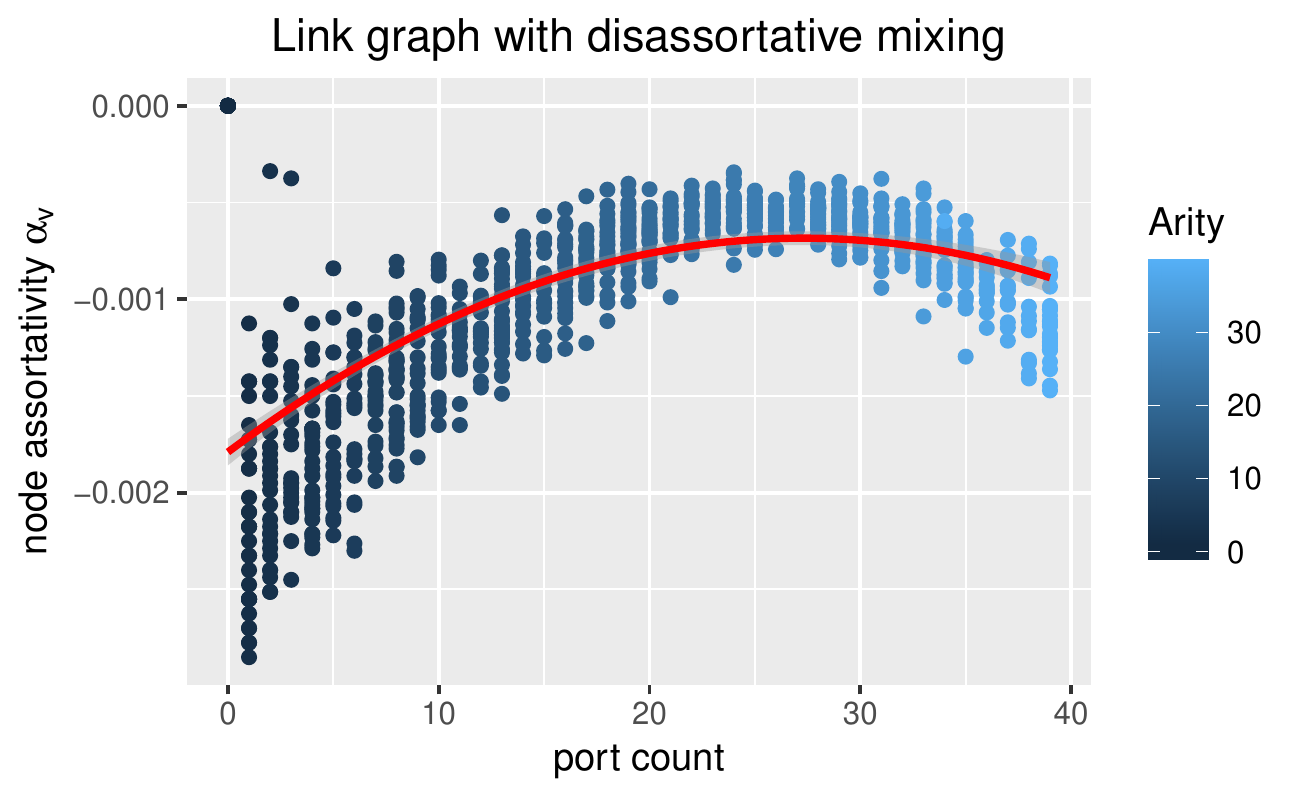}
		\caption{Node assortativity for a link graph with disassortative mixing.}
		\label{fig:node-assortativity-b}
	\end{subfigure}
	\caption{Node assortativity distribution of two link graphs ($N=1000$) generated under different configurations, namely, with an assortative (left) and disassortative (right) tendency. Each point represents a distinctive node. The x-axis denotes the number of ports of a node already occupied by a link, and the color denotes the associated arity of the node. The y-axis reflects the node assortativity of each node according to \Cref{eq:alternative-assortativity}.}
	\label{fig:analysis-linkgraph-assortativity}
\end{figure}

For the analysis we randomly generate two link graphs, one being assortative and one being disassortative (refer to the algorithm in \Cref{sec:link-graph-generation:maximal-degree-correlation}), each with $N=1000$ nodes and 40 controls with random arity $ar \in \{1,\ldots,40\}$ so that the node's arities obey a uniform distribution. Then, a node with arity $ar_i$ has the probability $\frac{1}{\abs{ar}}$ to be selected. The results are depicted in \Cref{fig:analysis-linkgraph-assortativity} and explained below. The values on the y-axis denote the assortativity of a node (\Cref{eq:alternative-assortativity}), the x-axis shows the actual port count of the node, and the color legend reflects the arity of a node.

The initial "arity sequence" derived by this selection process determines the frequency distribution of a node's ports available for linking. The algorithm can nearly connect every free port of a node with another node's port.
This is apparent from the color legend of \Cref{fig:analysis-linkgraph-assortativity}, which denotes the control's arity associated with a node. In other words, a counterexample of the current situation would be if a point colored light-blue (i.e., $ar > 30$) would be visible far left in the distribution (i.e., having high arity but low port count). The current case, however, illustrates that the link graph is nearly fully connected (no port is left unused).

\paragraph{Assortative Mixing}

The randomly generated assortative link graph is depicted in \Cref{fig:node-assortativity-a}, with mean $\mu_\alpha = 0.001$ and standard deviation $\sigma_\alpha = 0.00046$.
We have chosen $\lambda = \frac{2}{N} = 0.002$ assuming the link graph is perfectly assortative $r = 1$. The majority of nodes $58.2\%$ are slightly more assortative, and $40.6\%$ of the nodes being slightly more disassortative. The rest of the nodes ($1.2\%$) show a strong non-assortative linkage behavior.
To a great extent, \Cref{fig:node-assortativity-a} displays an equal node assortativity distribution with respect to the port count, meaning that hubs are generated, including nodes with similar arities.
A notable exception is that the left-most nodes exhibit both an assortative and disassortative mixing pattern. 
Moreover, the linear estimate exhibits an increasing trend leading to the remark that nodes with higher arity appear to be more assortative than the rest. Thus, it appears that high arity nodes tend to connect more quickly to other high arity nodes.

\paragraph{Disassortative Mixing}
The node assortativity distribution of the disassortative link graph is depicted in \Cref{fig:node-assortativity-b}. We chose $\lambda = 0$ assuming the link graph is perfectly disassortative $r = -1$, yielding $\alpha_v = -\hat{\delta}_v$, $\mu_\alpha = -0.001$ and $\sigma_\alpha = 0.00054$. Considering all disassortative nodes $\alpha_v < 0$, the fraction of nodes being slightly more assortative is $59.9\%$, and slightly more disassortative is $39.5\%$. The rest of the nodes ($0.6\%$) show a strong non-assortative linkage behavior. Moreover, the distribution in \Cref{fig:node-assortativity-b} exhibits very similar characteristics to the node assortativity distribution of the ER network ($N=1000$ nodes, $M=3000$ links) in \cite[Fig.~3]{thedchanamoorthy_node_2014}. Especially for the peripheral nodes, the distribution displays a more disassortative mixing compared to the core nodes around a port count of $25$. It can be observed from \Cref{fig:node-assortativity-b} that small arity nodes tend to connect more often to nodes with high arity nodes than vice versa. They have a much lower node assortativity value than the peripheral nodes on the right-hand side of the distribution. Because nodes with a small arity are removed early by the algorithm and high arity nodes remain available for linkage and share them between similar nodes at the end.

\section{Discussion and Conclusion}\label{sec:conclusion}

We have presented a bigraph generation algorithm with a preferential attachment feature for place graphs and two linking pattern variants regarding link graphs and analyzed some of their properties.
A main result of the work is that bigraphs can be efficiently generated from different standpoints: Our methods enable us to start either with the place graph or the link graph. The computational independence of both algorithms has the effect that one place graph can be created, and, based upon this, multiple different link graphs can be derived easily. Moreover, the individual algorithms allow the variation of several parameters, for example, the number of roots and nodes, choosing between a minimal or maximum number of connections between nodes; features that can provide good quality of results in a short period of time.

The results of the place graph generation algorithm in \Cref{sec:analysis-placegraphs} can be summarized as follows. The degree distribution follows a power-law distribution, and the distribution of nodes with positive arity obeys a binomial distribution. We verified that the preferential attachment feature is implemented by our algorithm, which is indicated by the long tail regarding the node degree distribution (see \Cref{fig:analysis-placegraph-degree-distribution}). 
It can be seen that nodes with many children accumulate more children faster (i.e., more children are nested under these). In other words, there are only a few branches in the place graph that are very deep, yet much more flat hierarchies exist.

Concerning the link graph generation algorithm, we presented two variants for the linkage behavior between nodes of a link graph. The first being termed \textit{Minimal Pairwise Port Linkage}. The algorithm yields a link graph where only two nodes at once are linked together. Regarding the second algorithm called \textit{Maximal Degree Correlation}, the link graph exhibits a fully connected pattern, where multiple connections between the same nodes are possible. It can be seen that the second approach is suitable for producing more complex link graphs. Owed to the fixed control's arity, the link graph generation algorithm does not achieve a distinctive assortative/non-assortative mixing pattern. This is due to the fact that nodes with a relatively small arity are removed early in the process and are thus not available anymore for linkage. However, we can observe when choosing the disassortative mixing variant that nodes with a high arity tend to connect to nodes with a low arity and small-arity nodes to high-arity nodes.

Unlike other random network algorithms, a link graph cannot create infinite links between nodes since the arity of a control represents a constraint that must be taken into account.
Thus, the node degree is prescribed because of the fixed arity of a control. This naturally leads to a more \textit{relative node assortativity} regardless of the original mixing pattern (assortative or disassortative) of the generated link graph.

\subsection{Reflections on the Application}

Our bigraph generation algorithm gives us the possibility to create synthetic graphs that are a useful resource within many domains of application. With this in mind, we can better address scalability problems of bigraphs by analyzing models focusing on properties such as the number and size of reaction rules and agents, also further by adding runtime variations to it for more sophisticated analyses.

For instance, fog-based applications based on bigraphical meta-models \cite{grzelak2019bigraph} can utilize synthetic models for benchmarking simulation methods.
With regards to a \textit{location model} such in \cite{grzelak2019bigraph}, we can test and simulate nearest neighbor or navigation queries within such a model. These queries are expressed as reaction rules. One can test whether certain reaction rules impact the number of synthesized states of a transition system of a BRS when these rules are applied (i.e., the bigraph matching and rewriting problem). An example of such a reaction rule might aim to locate specific nodes of a particular location where the location bigraph is randomly synthesized with our proposed method.

Moreover, we can more easily measure the performance of new model checking (see \cite{clarke_handbook_2018}) algorithms by generating many bigraph models under different constraints very conveniently.
When performing model checking, it would be helpful to build such models under different constraints for analyzing the effects of the state space explosion problem. One can investigate whether specific reaction rules affect state space explosion depending on the size or structure of an agent, for instance.

\subsection{Future Work}

We plan to include the notion of place and link sortings in our algorithms. 
Controls are equipped with so-called \textit{sorts} to ensure certain properties of a bigraph (e.g., a node assigned the control \textsf{Room} can not be nested under a node with the control \textsf{User}) (see \cite{milner_space_2009}). 
To implement this idea, our algorithms must be adapted in such a way that place and link sorts are preserved.
This feature may be necessary for creating more sophisticated bigraphs with regards to real-world ubiquitous computing applications where the semantic of structural containment plays a role.

We have chosen a purely algorithmic approach in our work. However, it is interesting to observe that we can make use of the bigraphical framework itself when creating random bigraphs. Such a pragmatic approach from a modeling standpoint was shown by Fernández et al. \cite{fernandez_labelled_2016}, where \textit{port graphs} are used. Nodes of port graphs "have explicit connection points called ports", similar to bigraphs, "to which edges are attached" \cite[p.~3]{fernandez_labelled_2016}. Different rewrite rules define the attachment behavior. We may adopt this in the future to evaluate its versatility.

\section*{Acknowledgment}
Funded by the German Research Foundation (DFG, Deutsche Forschungsgemeinschaft) as part of Germany's Excellence Strategy -- EXC 2050/1 -- Project ID 390696704 -- Cluster of Excellence "Centre for Tactile Internet with Human-in-the-Loop" (CeTI) of Technische Universität Dresden.

\section*{References}

\end{document}